\documentclass[preprint, showpacs, preprintnumbers, amsmath, amssymb, superscriptaddress, nofootinbib]{revtex4}
\usepackage{graphicx,color}
\usepackage{amsmath,amssymb,slashed}
\usepackage{url}
\usepackage{epstopdf}

\newcommand{\be}{\begin{equation}}
\newcommand{\ee}{\end{equation}}
\newcommand{\bea}{\begin{eqnarray}}
\newcommand{\eea}{\end{eqnarray}}

\newcommand{\crn}{\nonumber \\}

\newcommand{\la}{\lambda}

\newcommand{\om}{\omega}
\newcommand{\pa}{\partial}
\newcommand{\fr}{\frac}

\newcommand{\bc}{\begin{center}}
\newcommand{\ec}{\end{center}}

\newcommand{\ep}{\epsilon}

\newcommand{\ph}{\phi}

\newcommand {\ba}{\begin{array}}
\newcommand {\ea}{\end{array}}
\newcommand{\ben}{\begin{enumerate}}
\newcommand{\een}{\end{enumerate}}

\newcommand{\bre}{\allowdisplaybreaks}
\newcommand{\sla}[1]{\slash\!\!\!}

\usepackage{bm}
\usepackage{dcolumn}

\begin{document}

\title{ LFV decays in a 3-4-1 model with minimal inverse seesaw neutrinos}
\author{N.H.T. Nha}\email{nguyenhuathanhnha@vlu.edu.vn}
\affiliation{Subatomic Physics Research Group, Science and Technology Advanced Institute, Van Lang University, Ho Chi Minh City, Vietnam}
\affiliation{Faculty of Applied Technology, Van Lang School of Technology, Van Lang University, Ho Chi Minh City, Vietnam}

\author{L.T. Hue}\email{lethohue@vlu.edu.vn}
\affiliation{Subatomic Physics Research Group, Science and Technology Advanced Institute, Van Lang University, Ho Chi Minh City, Vietnam}
\affiliation{Faculty of Applied Technology, Van Lang School of Technology, Van Lang University, Ho Chi Minh City, Vietnam}
\author{L.T.T. Phuong}\email{lttphuong@agu.edu.vn}
\affiliation{{An Giang University,  An Giang, Vietnam}} 
\affiliation{{Vietnam National University, Ho Chi Minh City, Vietnam}} 

\author{T.T. Hong \footnote{corresponding author}}\email{tthong@agu.edu.vn}
\affiliation{{An Giang University,  An Giang, Vietnam}} 
\affiliation{{Vietnam National University, Ho Chi Minh City, Vietnam}}

\begin{abstract}
We investigate an extended 3-4-1 model consisting of a new singly charged Higgs boson, implementing the minimal inverse seesaw mechanism {to account for the current $(g-2)_{e,\mu}$ constraints} as well as the  lepton-flavor-violating decay rates of charged leptons, the  Standard Model-like Higgs boson, and the $Z$ boson, all consistent with current experimental data. Unlike the previously studied 3-4-1 realization,  the model considered here predicts strong correlations among these observables that can be tested in future experiments. In particular, the current upper bound on Br$(\tau \to \mu \gamma)$  imposes a stringent constraint compatible  with the $1\sigma$ experimental range of $(g-2)_{\mu}$, {corresponding to an allowed deviation of order} $10^{-9}$ from the SM prediction. The forthcoming experimental sensitivity to Br$(\tau \to \mu\gamma)$ will reduce this deviation to $5\times 10^{-10}$.

\end{abstract} 
\maketitle
\bre
 \section{Introduction}
 \label{sec:intro}
 The 3-4-1 model with right-handed neutrinos (341RHN) was introduced  in {Refs.~\cite{Foot:1994ym,Pisano:1994tf}} as a natural extension that  new right-handed neutrinos are assigned to the same left-handed lepton quadruplets.  For a complete  study of the highest possible gauge   symmetry in the electroweak sector \cite{Voloshin:1987qy}, various 3-4-1 extensions were introduced  with different electric charges of new exotic leptons \cite{Ponce:2003uu, Sanchez:2004uf, Ponce:2006vw, Sanchez:2008qv, Riazuddin:2008yx, Nam:2009tr,  Long:2016lmj, Palacio:2016mam, Djouala:2019hhn}. Interesting phenomenological implications, such as $(g-2)_{e_a}$ anomalies and lepton flavor violating (LFV) processes  involved with charged leptons, have been extensively studied \cite{Cogollo:2020nrc, Thao:2023gvs}.  It can be seen that 3-4-1 models can accommodate the latest experimental result for $(g-2)_{\mu}$ \cite{Muong-2:2023cdq, Muong-2:2025xyk},  in agreement with the most recent theoretical results calculated in the Standard Model (SM) framework \cite{Aliberti:2025beg}, in which the leading-order hadronicvacuum-polarization contribution derived from lattice-QCD calculations  has been taken into account \cite{Borsanyi:2020mff, RBC:2024fic, Djukanovic:2024cmq}. Consequently,  the updated SM prediction is compatible with  the experimental measurement; namely,  the corresponding deviation is given by~\cite{Aliberti:2025beg}:
 \begin{align}
 \label{eq:damu}
  \Delta a^{\mathrm{new}}_{\mu} &\equiv  a^{\mathrm{exp}}_{\mu} -a^{\mathrm{SM}}_{\mu} =\left(3.8\pm 6.3 \right) \times 10^{-10} .
 \end{align}  
  
  On the other hand, the original  3-4-1 models without exotic charged leptons predict small deviations, of $\Delta a_{\mu}\leq \mathcal{O}(10^{-11})$ \cite{Cogollo:2020nrc, Thao:2023gvs}, which are much smaller than the  $10^{-9}$ level associates with the current $1\sigma$ deviation and still within the reach of future experiments \cite{Hertzog:2025ssc, Athron:2025ets}. This behavior arises because  3-4-1models typically predict very heavy $SU(4)_L$ gauge bosons  at the  TeV scale,  resulting in suppressed one-loop contributions to $\Delta a_{\mu}$. In contrast,  various models featuring light neutral gauge and/or Higgs bosons can accommodate large upper value \cite{Li:2025myw, Aliberti:2025beg}.    The  experimental $a_e$ data  reported by different groups~\cite{Hanneke:2008tm, Parker:2018vye, Morel:2020dww, Fan:2022eto} indicate a possibility of large  deviations from the SM prediction of the  same order of magnitude, namely $|\Delta a^{\mathrm{NP}}_{e}|=\mathcal{O}(10^{-13})$, defined analogously to the muon case \cite{Aoyama:2012wj,  Laporta:2017okg, Aoyama:2017uqe,   Volkov:2019phy}.

Motivated by the {possible deviations in the $(g-2)_{e,\mu}$, which may be further clarified by future
	experiments,}  this work aims to investigate regions of the parameter space that  allow sizable $\Delta a_{e,\mu}$ in the 3-4-1 model supplemented by a  minimal number of  inverse seesaw (mISS) neutrinos and a singly charged Higgs boson. This model also predicts large LFV decay rates  within current and incoming experimental sensitivities.  In contrast to  the 341ISS  model discussed on  Ref. \cite{Thao:2023gvs}, we consider the 3-4-1 model consisting of the mISS neutrinos (341mISS), which may lead to strong correlations between LFV processes and $\Delta a_{e_a}$, particularly in the regions of the parameter space accommodating large $|\Delta a_{e,\mu}|$.  Studying these effects can help test the viability of the 3-4-1 models,  alongside many available  models beyond the SM {\cite{Han:2018znu, Endo:2019bcj, Li:2020dbg, DelleRose:2020oaa, Bigaran:2020jil, Botella:2020xzf, Bharadwaj:2021tgp, Han:2021gfu, Arbelaez:2020rbq, Chun:2020uzw, Chen:2020tfr, Dutta:2020scq, Wang:2022yhm, Hernandez:2021tii, Hernandez:2021xet, Li:2022zap, Botella:2022rte, Kriewald:2022erk, Barman:2021xeq, Dermisek:2022hgh, Chowdhury:2022jde, Chen:2023eof, Cao:2021lmj}}, which are motivated by various experimental results, including the {current constraints
	from} $(g-2)_{e,\mu}$. To the best of our knowledge, although various mechanisms, including the standard seesaw and ISS, for accommodating the observed active neutrino oscillation data within 3-4-1 frameworks have been  discussed  \cite{Palcu:2015ica, Palacio:2016mam}, the associated phenomenology, such as LFV decays and $(g-2)_{e_a}$ anomalies have not been investigated in detail. An exception in Ref. \cite{Thao:2023gvs}, which considered the standard ISS mechanism containing six ISS neutrinos. However, only correlations between $\Delta a_{\mu}$ and cLFV processes were  investigated numerically. In addition,  the 341ISS model contains more independent parameters  than the 341mISS model, including those from Yukawa couplings and ISS neutrino masses. Consequently, the 341ISS predictions  for  $\Delta a_{e_a}$ and cLFV decay rates are only weakly correlated with one another.

We will focus on the three  LFV decay processes that are currently being intensively searched for by experiments. Specifically, the latest experimental data and future sensitivities for the LFV processes- including decays of charged leptons  (cLFV) $e_b\to e_a \gamma$, the SM-like Higgs boson (LFV$h$), and the neutral gauge boson $Z$ (LFV$Z$)- are summarized in Table.~\ref{dataLFV}.
 \begin{table}[ht]  \footnotesize{
 		\begin{tabular}{cccc}
 			\hline
 			& Br & Latest experiment & Future sensitivity\\
 			\hline
 			\hline
 			& $\mathrm{Br}(\mu\rightarrow e\gamma)$ & $< 1.5\times 10^{-13}$ \cite{Venturini:2024keu, ParticleDataGroup:2024cfk, MEGII:2025gzr} & $<6\times 10^{-14}$ \cite{MEGII:2018kmf, Belle-II:2018jsg, MEGII:2025gzr} \\
 			cLFV & $\mathrm{Br}(\tau\rightarrow \mu\gamma)$ & $<4.2\times 10^{-8}$ \cite{ParticleDataGroup:2024cfk, Belle:2021ysv} & $< 6.9 \times 10^{-9}$ \cite{MEGII:2018kmf, Belle-II:2018jsg} \\
 			& $\mathrm{Br}(\tau\rightarrow e\gamma)$ & $<3.3\times 10^{-8}$ \cite{MEG:2016leq, Belle:2021ysv, BaBar:2009hkt, ParticleDataGroup:2024cfk} & $< 9.0 \times 10^{-9}$ \cite{MEGII:2018kmf, Belle-II:2018jsg} \\
 			\hline
 			& $\mathrm{Br}(h\rightarrow \mu^\pm e^\mp)$ & $<4.4\times 10^{-5} $\cite{ParticleDataGroup:2024cfk, CMS:2023pte} &  $\sim\mathcal{O}(10^{-5}) $ \cite{Qin:2017aju} \\
 			LFV$h$ & $\mathrm{Br}(h\rightarrow \tau^\pm \mu^\mp)$ & $<1.5\times 10^{-3} $\cite{CMS:2021rsq, ParticleDataGroup:2024cfk} &  $\sim\mathcal{O}(10^{-4}) $ \cite{Qin:2017aju, Barman:2022iwj, Aoki:2023wfb} \\
 			& $\mathrm{Br}(h\rightarrow \tau^\pm e^\mp)$ & $<2.0\times 10^{-3}$\cite{ParticleDataGroup:2024cfk, ATLAS:2023mvd} &  $\sim\mathcal{O}(10^{-4}) $ \cite{Qin:2017aju} \\
 			\hline
 			& $\mathrm{Br}(Z\rightarrow \mu^\pm e^\mp)$ & $<2.62\times 10^{-7} $\cite{ATLAS:2022uhq, ParticleDataGroup:2024cfk} & $\sim 7\times 10^{-8}$  (HL-LHC) and $10^{-10}$ (FCC-ee) \cite{ATLAS:2021bdj, Dam:2018rfz, FCC:2018byv} \\
 			LFV$Z$ & $\mathrm{Br}(Z\rightarrow \tau^\pm \mu^\mp)$ & $<6.5\times 10^{-6} $\cite{ATLAS:2021bdj, ParticleDataGroup:2024cfk} & $\sim 10^{-6}$  (HL-LHC)  and $10^{-9}$ (FCC-ee)  \cite{Dam:2018rfz, FCC:2018byv} \\
 			& $\mathrm{Br}(Z\rightarrow \tau^\pm e^\mp)$ & $<5.0\times 10^{-6}$\cite{ATLAS:2021bdj, ParticleDataGroup:2024cfk} & $\sim 10^{-6}$  (HL-LHC)  and $10^{-9}$ (FCC-ee) \cite{Dam:2018rfz, FCC:2018byv} \\
 			\hline
 		\end{tabular}
 		\caption{The latest and future  sensitivities on the LFV branching ratios (Br).    \label{dataLFV}} } 
 \end{table}
As shown in Table~\ref{dataLFV}, the projected experimental sensitivities for 2026 are expected to reach $\text{Br}(\mu \to e\gamma)<6\times 10^{-14}$, $\text{Br}(\tau \to e\gamma)<9.0\times 10^{-9}$, and $\text{Br}(\tau\to \mu \gamma)< 6.9 \times 10^{-9}$. These anticipated improvements will significantly enhance the sensitivity to cLFV processes. { We note that the pioneering works on LFV processes arising from active neutrino mixing were presented in Refs.~\cite{Petcov:1976ff,Bilenky:1977du,Cheng:1980tp}. In addition, several early studies of cLFV were based on spontaneously broken family symmetries with the emission of Goldstone bosons~\cite{Anselm:1985bp,Andreev:2006wh}, or involved leptoquarks~\cite{De:2024foq}; see also the review in Ref.~\cite{Lindner:2016bgg}, which discusses the close correlations between cLFV decays and $(g-2)_{e_a}$ anomalies. These developments suggest that a comprehensive and systematic analysis of cLFV signals is highly desirable, as it may help to maximize the sensitivity of future experiments probing interactions among different lepton generations.}

In this context, an important question arises: to what extent can sizable contributions to $\Delta a_{e_a}$, remain compatible with the increasingly stringent bounds on LFV decay rates carried out by incoming experiments?  In particular, it is crucial to determine whether large values of $\Delta a_{e_a}$  will be excluded by the forthcoming experimental limits on LFV processes and to identify the viable parameter space of the model.

The layout of this paper is as follows. Section \ref{341model} presents all ingredients of a 3-4-1 model needed to calculate the $(g-2)_{e_a}$ data and LFV decays. The Yukawa interactions in the lepton sector, as well as gauge boson masses, will be presented in this section. Then, we introduce the 341mISS model to accommodate the recent $(g-2)_{e_a}$ data, focusing on the allowed large values. Section \ref{1-loopcon} will show all of the one-loop contributions to the cFLV, LFV$h$, and LFV$Z$ decays with the corresponding Feynman rules and couplings. The detailed numerical results to determine the allowed regions of the parameter space that {accommodate the current
	experimental results for} $(g-2)_{e,\mu}$ and cLFV decays will be presented in Section \ref{sec_numer}. Finally, Section \ref{sec_con} summarizes important results and points out relevant comments. Additionally, Appendices \ref{app_gaugebosons}, \ref{potential}, and \ref{app_SMlikeHiggs} provide detailed derivations of the masses and mixings of all neutral gauge bosons, the Higgs potential and CP-even scalar sector, including the SM-like Higgs boson, as well as the relevant Higgs self-couplings in the 341mISS model.

\section{\label{341model} The 3-4-1 model with minimal inverse seesaw neutrinos }

Among various extension theories, the seesaw (type-I) mechanism explains the smallness of neutrino masses by introducing three right-handed (RH) neutrinos. It leads to an effective mass matrix for light Majorana neutrinos that is strongly suppressed relative to the electroweak scale when the RH neutrino mass matrix $M_R$ lies close to the typical grand unification scale. Namely, these neutrino masses are calculated by $m_\nu = -(m_D)^T(M_R)^{-1}m_D$ {\cite{Das:2018usr}}, and we know that if light neutrino masses are stabilized around sub-eV and a Dirac mass matrix $M_D$ comparable to the top-quark mass, then $M_R \sim 10^{14}$ GeV. Consequently, the direct testability of such conventional seesaw scenarios becomes highly challenging for current dectector. Besides, the inverse seesaw (ISS) framework is desirable, as it allows the new fermionic states to lie at the TeV scale while naturally generating light neutrino masses through a small lepton-number violating parameter {\cite{Isidori:2025rci}}. The mISS further reduces the particle content by introducing only two pairs of the SM gauge-singlet neutrinos \cite{Malinsky:2009df}, leading to a neutrino mass matrix with rank at most two. As a consequence, one light neutrino remains massless at tree level, which constitutes a distinctive and testable prediction of the mISS scenario. This minimal structure preserves compatibility with neutrino oscillation data while enhancing the predictability of LFV and precision observables.

 \subsection{\label{yukawa}Yukawa couplings and masses for fermions } 
 In this work, we will study the 3-4-1 model with heavy right-handed neutrinos and  new singly charged leptons assigned  to the three left-handed quadruplets  \cite{Sanchez:2004uf, Palacio:2016mam}.  The electric charge operator is defined as: $ Q= T_3 + \frac{1}{\sqrt{3}} T_8 -\frac{2}{\sqrt{6}} T_{15} + X \mathbb{I}.$
 The lepton sector  consists of three left-handed quadruplets, respectively right-handed singlets, {and the new singly charged Higgs boson $s^\pm$}, namely 
 \begin{align}
 \label{eq_lepton}	
 &	L_{a}   =   ( \nu'_a\, , e'_a \, , E'_a\, ,N'_a )_L^T \sim \left(1, 4,-\frac{1}{2} \right)\, ,\; a=1,2,3, \crn 
 &	e'_{a R},\; E'_{a R}  \sim   (1, 1, -1), N'_{aR}\sim   (1, 1, 0),\crn
 &{\nu'_{iR},  X'_{iR} \sim (1, 1,0), \; i =1,2, \, s^\pm \sim (1, 1, \pm1),}
 \end{align}
 where neutrino singlets will generated mISS neutrino masses. 
 
 The Higgs multiplets  and non-zero vacuum expectation values (VEV) of neutral components  needed for generating all fermion masses are given by: 
 \begin{align}
 \label{eq_Higgsmultiplet}
 \chi& = \left(%
 \chi_1^{0} \, ,
 \chi_2^{-} \, ,
 \chi_3^{-} \, ,
 \chi_4^{0}
 \right)^T \sim \left( 1, 4, -\frac{1}{2}\right), \; \langle \chi \rangle =    \left(%
 0 \, ,
 0\, ,
 0 \, ,
 \fr{v_\chi}{\sqrt{2}}
 \right)^T, 
 \crn  \phi &= \left(%
 \phi_1^{+}  \, ,
 \phi_2^{0} \, ,
 \phi_3^0 \, ,
 \phi_4^{+}
 \right)^T\, \sim \left( 1, 4,\frac{1}{2}\right), \; \langle \phi \rangle  =    \left(%
 0 \, ,
 0\, ,
 \fr{v_\om}{\sqrt{2}}\, ,
 0
 \right)^T, 
 \crn \rho& = \left(%
 \rho_1^{+}  \, ,
 \rho_2^{0} \, ,
 \rho_3^{0} \, ,
 \rho_4^{+}
 \right)^T \sim \left( 1, 4, \frac{1}{2}\right), \; \langle \rho  \rangle = \left(%
 0 \, ,
 \fr{v_1}{\sqrt{2}}\, ,
 0 \, ,
 0
 \right)^T, 
 \crn \eta &= \left(%
 \eta_1^{0}  \, ,
 \eta_2^- \, ,
 \eta_3^{-} \, ,
 \eta_4^{0}
 \right)^T\, \sim \left( 1, 4, -\frac{1}{2}\right), \;  \langle \eta \rangle  =    \left(%
 \fr{v_2}{\sqrt{2}} \, ,
 0\, ,
 0\, ,
 0
 \right)^T. 
 \end{align}
 Following the assignments of a new global $U(1)_{\mathcal{L}}$ symmetry defining a new general lepton number $\mathcal{L}$ relating to the normal lepton number. We assume here that The Yukawa Lagrangian respects this symmetry, unless a tiny violation from the mass term $\frac{1}{2} {\left( \mu_{X}\right)_{ij} \overline{X_{iR}}(X_{jR})^c} +\mathrm{ h. c.}$ with $i,j=1,2$ and $|\left({ \mu_{X}}\right)_{ij}|\leq 10^{-5}$ GeV. The lepton masses  are generated from the following Yukawa  interactions and mass term:
 \begin{align}
 \label{eq_Ylm}
 -\mathcal{L}_Y= & \;  y^{e}_{a b} \overline{L_{a} } \rho e'_{b R}   +y^{E}_{a b} \overline{L_{a}} \phi E'_{b R} + {y^{N}_{ab} \overline{L_{a} } \chi  N'_{b R}} +   y^{\nu}_{a i} \overline{L_{a} } \eta  \nu'_{i R} + \frac{1}{2}{ \left( \mu_{X}\right)_{ij} \overline{X_{iR}}(X_{jR})^c}
 +\mathrm{ h. c.}. 
 \end{align}
From the last two terms in Eq.~\eqref{eq_Ylm} and via the ISS mechanism, we rewrite the Lagrangian Yukawa interactions for neutrinos sector as Eq.~\eqref{eq_mISS}, and they will be expanded in detail below.
\begin{align}
-\mathcal{L}_{Y,\nu}=&{  \overline{\nu_R} y^{\nu}\eta^{\dagger}L  + \overline{\nu_{R}}M_{R}(X_{R})^c   +\frac{1}{2} \overline{X_{R}}\mu_{X}(X_{R})^c  + \overline{X^c}  y^se'_R s^+ + \mathrm{h.c.}},
\label{eq_mISS}
\end{align}
 The model consists of quark multiplets  that must be arranged to cancel gauge anomalies, see for example a discussion in Ref. \cite{Long:2016lmj}. It can be seen that the quark masses can be constructed to satisfy the recent experimental data.   This sector is irrelevant to our work. 
The mass terms of all leptons are:
 \begin{align} \label{eq_chargedmass}
(M_{e})_{a b}= y^e_{a b} \fr{v_1}{\sqrt{2}}\,, \; (M_{E })_{a b} = y^E_{a b} \fr{v_\om}{\sqrt{2}},{(m_{D})^*_{a i}= {y^\nu_{a i}} \fr{v_2}{\sqrt{2}} } \,,\;  {(M_{N })_{a b} = y^N_{a b} \fr{v_\chi}{\sqrt{2}}.}
 \end{align}
{where, $y^e_{a b}, y^E_{a b}, y^N_{a i}$ are $3\times3$ matrices, while $y^\nu_{a i}, $ is a $3\times2$ matrix generating active neutrino masses through the minimal seesaw mechanism. The relevant Dirac neutrino mass matrix is $M_{\nu}$.}  However, these tiny masses do not affect significantly the one-loop contributions to $a_{e_a}$. 
 
  We now focus on the lepton sector in the Yukawa part of Eq. \eqref{eq_Ylm}.  The normal lepton mass matrix $M_e$ given in Eq.~\eqref{eq_chargedmass} is assumed to be diagonal for simplicity. As a result, the flavor basis of the charged leptons $e'_a$ coincides with the mass basis $e_{aL,R}\equiv e'_{aL,R}$, namely 
 \begin{equation}\label{eq_SMmasse}
 m_{e_a}= y^e_{ab}\delta_{ab} \frac{v_1}{\sqrt{2}} \Rightarrow y^e_{ab}=\delta_{ab}\frac{\sqrt{2}m_{e_a}}{v_1}, 
 \end{equation}
where $\delta_{ab}= 1 (a=b), \text{and}\, \delta_{ab}= 0 (a\neq  b)$. Three other base $l'_{L,R}\equiv (l'_1, l'_2, l'_3)^T_{L,R}$ with {$l=E,N$}  are transformed generally into the corresponding mass bases $l'_{L,R}$ through the following relations:
 \begin{align}
 \label{eq_Lmixing}
 {U^{l\dagger}_L M_lU^{l}_R} &=\hat{M}_{l}=\mathrm{diag}(m_{l_1},m_{l_2},m_{l_3}),\; l'_{L,R} =U^{l}_{L,R}l_{L,R}. 
 \end{align}
 
 \subsection{\label{gaugeboson} Gauge bosons}
 Gauge boson masses arise from the covariant kinetic term of Higgs multiplets, namely 
 \begin{equation}\label{eq_LkH}
 L_{\mathrm{Higgs}} = \sum_{H}^4 \left(D^\mu\langle H \rangle\right)^\dag D_\mu \langle H \rangle ,
 \end{equation}
 where $H=\chi, \phi,\eta,\rho$. The covariant derivative is defined as
 \begin{align} \label{eq_defDmu}
 D_\mu & = \pa_\mu - i g\sum_{a=1}^{15} W_{ a\mu} T_a  - i g_X X B''_\mu T_{16}
 \equiv  \pa_\mu - i g P_\mu^{CC} - i g P_\mu^{NC}\, ,
 \end{align}
{where $g$ and $g_X$ are the gauge couplings, while $W_{a \mu}$ and $B''_\mu$ are the gauge fields associated with the $SU(4)_L$ and $U(1)_X$ gauge groups, respectively. The two terms $P_\mu^{NC}$ and $P_\mu^{CC}$ correspond to the neutral and non-Hermitian currents \cite{Long:2016lmj}.}
 For quadruplet, $T_{16} = \fr{1}{2\sqrt{2}} \textrm{diag} (1,1,1,1)$, and 
 \begin{equation}
 \label{eq_defPCC}
 P_\mu^{CC} = \fr{1}{\sqrt{2}}\left(%
 \begin{array}{cccc}
 0 & W^+ &W_{13}^{+} & W_{14}^{0}\\
 W^-  & 0 & W_{23}^{0}  &W_{24}^{-}\\
 W_{13}^{-} & W_{23}^{0*} & 0 &W_{34}^{-} \\
 W_{14}^{0*}&W_{24}^{+}&W_{34}^{+}&0
 \end{array}\,
 \right)_\mu ,\;	P_{\mu}^{NC}= \frac{1}{2}\text{diag}\left(W^+_{3815}, W^-_{3815}, W_{815}, W_{15}\right), \crn
 \end{equation}
 where  {$t_X\equiv g_X/g=2\sqrt{2}s_W/\sqrt{1-2s_W^2}$}, $\sqrt{2}\, W^\mu_{ij} \equiv W^\mu_i - i W^\mu_j$ with $i<j$,  $W^\pm_{3815} = \pm W_{3\mu} + \frac{W_{8\mu}}{\sqrt{3}} + \frac{W_{15\mu}}{\sqrt{6}} + \frac{Xt_XB_{\mu}^{''}}{\sqrt{2}},\; W_{815} = - \frac{2W_{8\mu}}{\sqrt{3}}+\frac{W_{15\mu}}{\sqrt{6}} + \frac{Xt_XB_{\mu}^{''}}{\sqrt{2}}$, and $W_{15} = -\frac{3W_{15\mu}}{\sqrt{6}} + \frac{Xt_XB_{\mu}^{''}}{\sqrt{2}}$.
 The upper subscripts label the electric charges of gauge bosons. For completeness, we briefly comment on the Higgs and gauge boson sectors. The masses and mixings of the neutral gauge bosons are presented in Appendix~\ref{app_gaugebosons}; see Ref.~\cite{Long:2016lmj} for detailed calculations. The Higgs potential, as well as the singly charged Higgs bosons, is discussed in detail in Appendix~\ref{potential}, following Ref.~\cite{Thao:2023gvs}. The properties of the SM-like Higgs boson are presented in Appendix~\ref{app_SMlikeHiggs}.
 	
 The prospects for LHC searches for the extra neutral gauge bosons  $Z_3$ and $Z_4$ predicted by  3-4-1 models were discussed in Ref. \cite{Lee:2014kna}, suggesting a compelling way to distinguish between 3-3-1 and 3-4-1 gauge extensions. In particular, the searches for exotic neutral gauge bosons $Z_3$ and $Z_4$  through the processes $q\bar{q},e^+e^-\to Z' (Z'')\to \bar{f}f$ are of considerable interest in BSM scenarios with extended electroweak gauge symmetries.  In particular, the 3-3-1 and 3-4-1 models predict the existence of one ($Z_3$) and two ($Z_{3,4}$) neutral gauge bosons, respectively. Because the $V-A$ gauge couplings $g^{X}_{V,A}$ ($Z=Z_3,Z_4$) of these bosons to SM fermions, namely leptons and quarks, encode the structure of the underlying electroweak gauge symmetry and therefore determine their decay rates and production cross sections at colliders. Consequently, if these neutral gauge bosons are observed at future colliders, the corresponding experimental data can be used to determine their couplings to fermions, which are essential for  identifying the underlying 3-3-1 or 3-4-1 model. 

 \subsection{\label{sec_lSSmodel} The 3-4-1 with mISS neutrinos}
 
In the Eq.~\eqref{eq_mISS}, $y^{\nu}$ is $3\times2$ matrices, while $M_R$, $\mu_X$, and $y^s$ are $2\times2$ matrices. Notations for flavor states of active left-handed neutrinos are $ \nu_{L}=(\nu'_{1},\nu'_{2},\nu'_{3})^T_L$ and  $ \nu_R=(\nu_{1}, \nu_{2})^T_R$,  $ X_R=( X_{1}, X_{2})^T_R$. The $7\times7$ neutrino mass matrix $M^\nu$ in the basis $(\nu_L, \nu_R, X_R)$ is derived by rewritten these terms in Eq.~\eqref{eq_mISS} in the following ISS form
  \begin{align}
 - \mathcal{L}^{\nu}_{\mathrm{mass}}
 &\equiv\frac{1}{2}\left(\overline{(\nu_L)^c},\; \overline{\nu_R}, \; \overline{X_R}\right)
 \mathcal{M}^{\nu} \begin{pmatrix}
 \nu_L,
 \left( \nu_R\right)^c,
 \left( X_R\right)^c
 \end{pmatrix}^T 
 + \mathrm{h.c.},
 \crn  
 \mathcal{M}^{\nu}&= 
 \left(
 \begin{array}{cc}
 \mathcal{O}_{3\times3} & {M_D^T} \\
 {M_D} & M_{N} \\
 \end{array}
 \right), \;  {M_D^T= \begin{pmatrix}  m_D^* &  \mathcal{O}_{3\times2}
 \end{pmatrix}}, 
 \;  M_N=\left(
 \begin{array}{cc}
 \mathcal{O}_{2\times2}& M^T_R \\
 M_R & \mu_X \\
 \end{array}
 \right), 
 \label{eq_L0ISSnumass}	
 \end{align}
 where $\mathcal{O}_{3\times3}$, $\mathcal{O}_{3\times2}$, and $\mathcal{O}_{2\times2}$ is zero matrices, the $\mathcal{M}^\nu$ has form similar in Ref.~\cite{ParticleDataGroup:2024cfk},  and {$m_D^*=y^{\nu}\times v_2/\sqrt{2}$.} In the limit $\mu_X\to \mathcal{O}_{2\times2}$, the rank $\mathcal{M}^\nu$ reduces from $6$ to $4$, which will leave three light neutrinos massless, which exactly matches the SM. The most significant aspect of mISS is that it predicts that the lightest neutrino is exactly massless because the Dirac mass matrix $m_D$ is $3\times2$ matrix have rank $2$, which differs from the original ISS. Hence, $m_{n_1(n_3)} = 0$ for normal order (NO) or inverse order (IO) scheme.

The analytic form of the Dirac mass matrix was chosen generally following Ref. \cite{Casas:2001sr}. The total unitary mixing matrix $U^\nu$  is defined  as follows 
 \begin{align}
 U^{\nu T}\mathcal{M}^{\nu}U^{\nu}=\hat{\mathcal{M}}^{\nu}=\mathrm{diag}(m_{n_1},m_{n_2},m_{n_3}, m_{n_4},..., m_{n_{7}})\equiv \mathrm{diag} \left( \hat{m}_{\nu},\; \hat{M}_N\right), \label{eq_dUnu}	
 \end{align}
 where   $m_{n_m}$ ($m=1,2,...,7$) are  eigenvalues of the $7$ mass eigenstates $n_{mL}$,  including three  light active neutrinos $n_{aL}$ ($a=1,2,3$) with mass matrix $\hat{m}_{\nu}$ and four other heavy neutrinos with mass matrix $\hat{M}_N$. 
 The relation between the flavor  and mass eigenstates are
 \begin{equation}
 \begin{pmatrix}
 \nu_L,
 \left( \nu_R\right)^c,
 \left( X_R\right)^c	
 \end{pmatrix}^T =U^{\nu} n_L  \; \mathrm{and} \; \begin{pmatrix}
 \left( \nu_L\right)^c,
 \nu_R,
 X_R	
 \end{pmatrix}^T =U^{\nu*}  n_R, \label{eq_Nutrans}
 \end{equation}
 where $n_L\equiv(n_{1L},n_{2L},...,n_{7L})^T$ and $n_R=\left(n_L\right)^c$. 
 %
 The neutrino mixing matrix is parameterised in the following form
 \begin{equation} 
 U^{\nu}= \left(
 \begin{array}{cc}
 \left(	\mathbb{I}_3-\frac{1}{2}RR^{\dagger} \right) U_{\mathrm{PMNS}} & RV_4 \\
 -R^\dagger U_{\mathrm{PMNS}} & \left(\mathbb{I}_4 -\frac{1}{2}R^{\dagger} R\right)V_4 \\
 \end{array}
 \right),  
 \label{eq_Unu0}	
 \end{equation}
 where   $U_{\mathrm{PMNS}}$  is the  $3\times3$  Pontecorvo-Maki-Nakagawa-Sakata (PMNS) matrix \cite{Pontecorvo:1957cp, Maki:1962mu},  $V_4$ is a $4\times 4$ unitary matrix,  and  $R$ is a $3\times 4$ matrix satisfying $|R_{ax}|\ll1$ for all $a=1,2,3$ and $x=1,2,3,4$.
 In the ISS framework we considered  here, $m_D$ is parameterized in terms of many free parameters, hence it is enough to choose a simple form of  $\mu_X=\text{diag}(\mu_1,\mu_2)=\mu_0 \mathbb{I}_2$, and    $M_R=\hat{M}_R= M_0\mathbb{I}_2$ {as similar with ISS neutrinos in Refs.~\cite{Arganda:2014dta, Thao:2017qtn}.}
 The limit above is similar ISS scenario with $\mathbb{I}_2 \to I_3$, and $m_D=M_0\sqrt{\hat{x}_\nu}U^\dagger_\mathrm{PMNS}$ {discussed on Ref. \cite{Thao:2023gvs}}. In the framework, we will investigate the general heavy neutrinos' mass not degeneracy, namely, $M_R=\hat{M}_ R=\text{diag}(M_1, M_2)$. {The formulas of  $m_D$  and mixing parameters  are  parameterized as followed \cite{Casas:2001sr}
 \begin{equation}
 {m_D^{\dagger}}= \hat{M}_RU_{\zeta} \sqrt{\hat{x}_\nu}  U^{\dagger}_{\mathrm{PMNS}},\;  R  \simeq \left(0,\;  U_{\mathrm{PMNS}}\hat{x}_\nu^{1/2} U_{\zeta}^{\dagger} \right), \; \hat{x}_\nu\equiv \frac{\hat{m}_\nu}{\mu_0}, \label{eq_mDiss}
 \end{equation}
 where  max$[\left|\left(\hat{x}_{\nu}\right)_{aa} \right|]\ll1$ for all $a=1,2,3$ and the matrix $U_{\zeta}$ corresponding to the MISS framework in the two NO and IO scenarios are \cite{Hong:2024yhk} 
 \begin{equation}\label{eq:zeta}
 U_{\zeta}(\mathrm{NO})= \begin{pmatrix}
 0&c_{\zeta} &-s_{\zeta} \\
0 & s_{\zeta}& c_{\zeta}
 \end{pmatrix}, \;  U_{\zeta}(\mathrm{IO})= \begin{pmatrix}
c_{\zeta} &-s_{\zeta} &0 \\
 s_{\zeta}& c_{\zeta} &0
 \end{pmatrix}.
  \end{equation}
}
 The ISS conditions $|\hat{m}_{\nu}|\ll |\mu_0|\ll |m_D|\ll M_1, M_2$ so that $\dfrac{\sqrt{\mu_0 \hat{m}_{\nu}}}{M_1, M_2}\simeq0$,  the mixing matrix and Majorana mass term are 
 \begin{align}
 \hat{M}_N= \left(\begin{matrix}
 \hat{M}_R	& 0 \\ 
 0	& \hat{M}_R
 \end{matrix} \right), 
 \;  V_4\simeq \dfrac{1}{\sqrt{2}}
 \left(\begin{matrix}
 -i\mathbb{I}_2 	& \mathbb{I}_2   \\ 
 i\mathbb{I}_2 		& \mathbb{I}_2 
 \end{matrix} \right), \label{eq_UNiss}	
 \end{align}
 which give $V_4^*\hat{M}_NV_4^{\dagger} \simeq M_{N}$, i.e.,  $m_4=m_6=M_1$, and $m_5=m_7=M_2$.
 
We have rewriten the Eq.~\eqref{eq_Unu0} as below
 {
 \begin{equation} 
 U^{\nu}= \left(
 \begin{array}{ccc}
 U_{\mathrm{PMNS}}\left(\mathbb{I}_3-\frac{1}{2}\hat{x}_\nu\right) & \frac{iU_{\mathrm{PMNS}} \hat{x}_\nu^{1/2}U_{\zeta}^{\dagger}}{\sqrt{2}} & \frac{U_{\mathrm{PMNS}} \hat{x}_\nu^{1/2}U_{\zeta}^{\dagger}}{\sqrt{2}} \\
 0 & \frac{-i\mathbb{I}_2}{\sqrt{2}} & \frac{\mathbb{I}_2}{\sqrt{2}}\\
 -U_{\zeta} \hat{x}_\nu^{1/2} &  \frac{i}{\sqrt{2}}\left(	\mathbb{I}_2-\frac{U_{\zeta}\hat{x}_\nu U_{\zeta}^{\dagger}}{2} \right) & \frac{1}{\sqrt{2}}\left(	\mathbb{I}_2-\frac{U_{\zeta}\hat{x}_\nu U_{\zeta}^{\dagger}}{2}\right) \\
 \end{array}
 \right). 
 \label{eq_Unu0t}	
 \end{equation}
 }
The lepton mixing matrix Eq.~\eqref{eq_Unu0t} generates the LFV couplings as {the} main sources of LFV processes, which appear in the kinetic Lagrangian of leptons and {the} Yukawa Lagrangian.  The lepton kinetic terms are:
  \begin{align}
  \label{eq:lkLep}
  \nonumber\mathcal{L}^{\ell \ell V}=&\sum_{a=1}^3 \left[  i \overline{L_{a L}} \gamma^\mu D_\mu L_{a L} +  {i \overline{N'_{a R}} \gamma^\mu \partial_\mu N'_{a R} + i \overline{E'_{a R}} \gamma^\mu D_\mu E'_{a R}} \right] \crn
  =& \sum_{a=1}^3 i \begin{pmatrix}
  \overline{\nu_{aL}^{'}}&\overline{e_{aL}^{'}}&\overline{E_{aL}^{'}}&\overline{N_{aL}^{'}}
  \end{pmatrix} \gamma^\mu \left[\pa_\mu-ig \left(P_{\mu}^{NC}+P_{\mu}^{CC} \right) \right]
  \begin{pmatrix}
  \nu_{aL}^{'}, 
  e_{aL}^{'},
  E_{aL}^{'},
  N_{aL}^{'}
  \end{pmatrix}^T + .... \crn
  =&\sum_{a=1}^3 \frac{g}{\sqrt{2}}\left[ \sum_{m=1}^7U_{am}^{\nu*}\overline{n_{m}}W_{\mu}^{+}\gamma^{\mu}P_{L}e_{a}+ \sum_{b=1}^3\left( U_{ab}^{N*}\overline{N_{b}}\gamma^{\mu}P_{L}W_{24\mu}^{+}e_{a} + U_{ab}^{E}\overline{e_{a}}\gamma^{\mu}P_LW_{23\mu}^{0}E_{b}\right)+ \text{h.c}\right] 
  \crn & + .... , 
  \end{align}
  where we list here only couplings give one-loop contributions to $(g-2)_{e_a}$ and cLFV decays. We can see that the $SU(4)_L$ particles such as $W_{24}$ and $W^0_{23}$,  generate one-loop contribution to $\Delta a_{e_a}$.  In addition, LFV couplings may also appear if $U^{N}, U^E\neq I_3$. On the other hand,  previous results \cite{Thao:2023gvs} show that 1-loop contributions from $SU(4)_L$ particles to $\Delta{a}_{\mu}$  can reach only maximal values of $\mathcal{O}(10^{-11})$, {which} under recent experimental constraint of heavy neutral gauge boson searches. We confirm again that they also result in small LFV effects. Therefore, in the numerical investigation, we will ignore the effects arising from $SU(4)_L$ gauge boson exchanges. 
  
 The Yukawa Lagrangian generating LFV couplings involving Higgs bosons are as follows. The contributions from ISS neutrino couplings with singly charged
 Higgs bosons derive from use physical states in Eqs.~(\ref{eq_rho1eta2}, \ref{ph_stateh}) are listed in the Lagrangian below
 \begin{align}
\label{Dnhat}
-\mathcal{L}_Y =&\sum_{a=1}^3\left[ \frac{\sqrt{2}m_{e_a}}{v c_{\beta}}\left(\overline{e^{'}_{aL}}\rho_2^0 + \overline{\nu_{aL}^{'}}\rho_{1}^{+}\right) e_{aR} +\sum_{i=1}^2\left(\frac{{(m_{D})^*_{ai}}\sqrt{2}}{v s_{\beta}}\left(\overline{\nu_{aL}^{'}}\eta_{1}^{0} +\overline{e_{aL}^{'}}\eta_{2}^{-}\right)\nu_{iR}^{'} + {y_{ia}^{s}}\overline{X_{iR}^{c}}e_{aR}s^{+} \right)\right]
\crn& +\text{h.c.}\crn
=&{\sum_{a=1}^3\left(\frac{gm_{e_a}s_{\alpha_0} }{2m_{W}c_{\beta}}h\overline{e_a}e_a \right)+ h \sum_{m,n=1}^7  \left[ \frac{gc_{\alpha_0}}{2m_{W}s_{\beta}}\overline{n_{m}}\left(\lambda'^*_{nm}P_R +\lambda'_{mn}P_L\right)n_n\right]}\crn
&  {+ \frac{g}{\sqrt{2}m_{W}}\sum_{a=1}^3\sum_{m=1}^7\sum_{j=1}^2\left[ \overline{n_{m}}\left( \lambda_{am}^{L,j} P_{L} +  \lambda_{am}^{R,j}P_{R} \right) e_ah_{j}^{+} +\mathrm{h.c.} \right]+....,} 
\end{align} 
where 
\begin{align}
\lambda_{am}^{L,1} =& t^{-1}_{\beta}c_{\alpha_\pm} \sum_{i=1}^2{(m_{D})^*_{ai}}{U_{(i+3)m}^{\nu}}, \; 
\lambda_{am}^{R,1}= t_{\beta}c_{\alpha_\pm}m_{e_a}U_{am}^{\nu*} -\frac{\sqrt{2}m_{W}s_{\alpha_\pm}}{g}\sum_{i=1}^2{y_{ia}^{s}}U_{(i+5)m}^{\nu*},\crn
\lambda_{am}^{L,2} =& t^{-1}_{\beta}s_{\alpha_\pm}  \sum_{i=1}^2{(m_{D})^*_{ai}}{U_{(i+3)m}^{\nu}}, \; 
\lambda_{am}^{R,2}= t_{\beta}s_{\alpha_\pm}m_{e_a}U_{am}^{\nu*} +\frac{\sqrt{2}m_{W}s_{\alpha_\pm}}{g} \sum_{i=1}^2{y_{ia}^{s}}U_{(i+5)m}^{\nu*},
\crn {\lambda'_{mn}} =& \sum_{i=1}^2 \sum_{a=1}^3 {(m_{D})^*_{ai}}U_{(i+3)m}^{\nu}U_{an}^{\nu}. 
\end{align}
In practice, the Feynman rule for {$h\overline{n}n$} couplings is written  in {the symmetric form \cite{Pilaftsis:1992st, Korner:1992zk, Dreiner:2008tw}}  as $ \sum_{m, n=1}^7 h\overline{n_{m}}\left(\lambda'_{mn}P_L +\lambda'^*_{nm}P_R\right)n_n= \frac{1}{2}\sum_{m, n=1}^7 h\overline{n_{m}}\left(\lambda_{mn}P_L +\lambda_{mn}^*P_R\right)n_n$ with
$$\lambda_{mn} =\lambda_{nm}= \sum_{c=1}^3\left(U_{cm}^{\nu} m_{n_m}U_{cn}^{\nu*} + U_{cn}^{\nu} m_{n_n}U_{cm}^{\nu*}\right). $$
 Here, only couplings of singly charged Higgs boson giving one-loop contributions to $(g-2)_{e_a}$ and cLFV decays. The first line are couplings relevant to one-loop contributions to LFV$h$ decays.

\section{\label{1-loopcon} Analytic formulas for one-loop contributions} 
\subsection{$h\to e_b^\pm e_a^\mp$ decays}

The couplings of scalar Higgs $h$ with gauge bosons from the covariant kinetic term of Higgs multiplets in Eq.~\eqref{eq_LkH}. Similarly, to find the couplings between scalar bosons, two scalars and a charged gauge boson, we expand the Higgs potential in the expression Eq.~(\ref{eq_Vh}), noting that only the third-order interaction vertices will be retained. The results for relevant coupling factors are presented in Table.~\ref{cp_higgs}, and we note that the notations for the Feynman rules corresponding to vertices and coupling factors in all the Tables below are used similarly to those in Ref.~\cite{Hue:2024rij},
\begin{table}[ht]
	\centering 
	\renewcommand{\arraystretch}{1.2}
	\begin{tabular}{cccc}
		\hline
		Vertex & factor & Vertex & factor\\
		\hline
		\hline
		$g_{hW^+ W^-}$	& $gm_W \text{sin}(\alpha_0 +\beta)$	& $g_{hW_{24}^+ W_{24}^-}$	& $gm_W s_{\alpha_0}c_{\beta}$	\
\\
		$g_{hh_{1}^{+} W^-}$ & $\frac{g}{2}c_{\alpha_\pm} \text{cos}(\alpha_0 +\beta)$ & $g_{hh_{1}^{-} W^+}$ & $-\frac{g}{2}c_{\alpha_\pm} \text{cos}(\alpha_0 +\beta)$	
\\
		$g_{hh_{2}^{+} W^-}$ & $\frac{g}{2}s_{\alpha_\pm} \text{cos}(\alpha_0 +\beta)$ & $g_{hh_{2}^{-} W^+}$ & $-\frac{g}{2}s_{\alpha_\pm} \text{cos}(\alpha_0 +\beta)$	
\\
$g_{hh_i^+h_j^-}$	& $\lambda_{hij}$ & $g_{hh_{i}^\pm W_{24}^\mp}$ & 0 
\\
		\hline			
	\end{tabular}
	\caption{The coupling factors of the SM-like Higgs boson $h$ to singly charged Higgs and gauge bosons in the 341mISS model, in which $i,j=1,2$. \label{cp_higgs} }
\end{table}
where $m_W = \frac{gv}{2}$, $\delta \equiv \pi/2-\alpha_0-\beta$, and $\lambda_{hij}$ are detailed presented in Eq.~\eqref{factor_hij}. Note that although the coupling factors $g_{hW^{+}_{24}W^{-}_{24}}$ may be as large as $g_{hW^{+}W^{-}}$, the respective two one-loop contributions relevant to them differ qualitatively from each other by a factor of $m_W/m_{W_{24}}$, therefore the Feynman diagrams with the $W_{24}$ exchange give small one-loop contribution to $\Delta a_{e_a}$. We ignore this one-loop contribution in the numerical investigation.

Matching the SM result of  the couplings $hWW$ and $hZZ$ result in that  $\sin(\alpha_0+\beta)=1$, leading to the limit that  {$\alpha_0+\beta \equiv \pi/2 -\delta \to \pi/2$}, i.e.  $\delta\to0$. In numerical investigation, we choose the dependence that $\alpha_0=\pi/2-\beta-\delta$, while $\beta$ and $\delta$ are free parameters constrained by both theoretical and experimental results.

\begin{figure}[ht]
	\includegraphics[width=16.5cm]{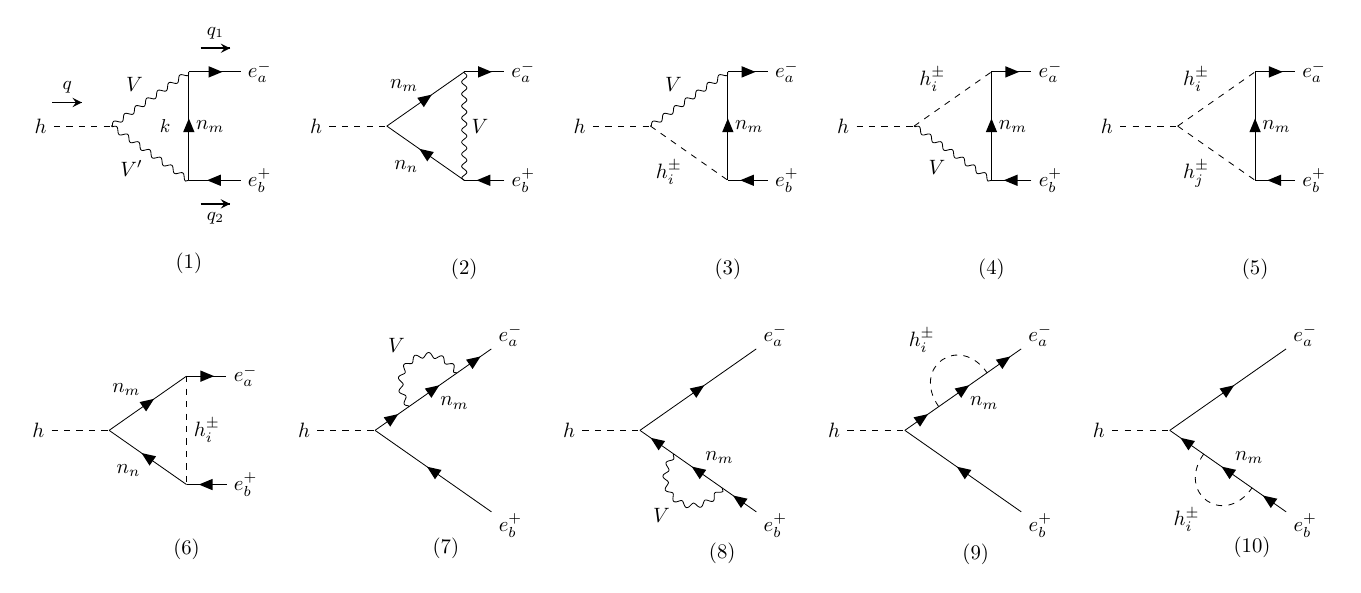}
	\caption{Feynman diagrams giving  one-loop contributions  to the decay $h\rightarrow e^-_ae^+_b$ in the unitary gauge, where $h_i^\pm=h_1^\pm, h_2^\pm$, and  $V=W^{\pm}$.}\label{fig_LFVh}
\end{figure}

In the unitary gauge, the Feynman diagrams for one-loop contributions to LFV$h$ decays are shown in Fig. \ref{fig_LFVh}   \cite{Arganda:2004bz, Arganda:2014dta, Nguyen:2018rlb, Hong:2022xjg}.  
The effective Lagrangian of the  decays  is
$$ \mathcal{L}^{\mathrm{LFV}h}= h \overline{e_a} \left[\Delta_{(ab)L} P_L  +\Delta_{(ab)R}  P_R \right]  e_b + {\mathrm{h.c.}},$$
where the scalar factors $\Delta_{(ab)L,R}$ are considered as  one-loop contributions in this work.  In the limit   $m^2_{h}\gg m^2_{e_a,e_b}$ with $m_{e_a,e_b}$ being charged lepton masses, 
the partial width  is
\begin{equation}
	\Gamma (h\rightarrow e_ae_b)\equiv\Gamma (h \rightarrow e_a^{-} e_b^{+})+\Gamma (h\rightarrow e_a^{+} e_b^{-})
	\simeq   \fr{ m_{h} }{8\pi }\left(\vert \Delta_{(ab)L}\vert^2+\vert \Delta_{(ab)R}\vert^2\right){.} 
	\label{LFVwidth}
\end{equation}
 The on-shell conditions for external particles are $p^2_{1,2}=m_{e_a,e_b}^2,$  and $ q^2 \equiv( p_1+p_2)^2=m^2_{h}$. The corresponding branching ratio is  Br$(h \rightarrow e_ae_b)= \Gamma (h\rightarrow e_ae_b)/\Gamma^{\mathrm{total}}_{h},$ where $\Gamma^{\mathrm{total}}_{h}\simeq 4.1\times 10^{-3}$ GeV \cite{Denner:2011mq}.  The $\Delta_{(ab)L,R}$ can be written as follows 

\begin{align}
\Delta_{(ab)L,R} =& \sum_{k=1,7,8} \sum_{m=1}^7\Delta^{(k)n_mWW}_{(ab)L,R}   + \sum_{m,n=1}^7\Delta^{{(2)}Wn_mn_n}_{(ab)L,R} +\sum^2_{i=1}\sum_{m=1}^7\left(\Delta^{(3)n_mWh_i^\pm }_{(ab)L,R} +\Delta^{(4)n_m h_i^\pm W}_{(ab)L,R} \right)
\crn &+ \sum_{i,j=1}^2\sum_{m=1}^7\sum_{k=5,9,10} \Delta^{(k)n_mh_i^\pm h_j^\mp}_{(ab)L,R} + \sum_{i=1}^2\sum_{m,n=1}^7 \Delta^{(6)h_i^\pm n_mn_n}_{(ab)L,R}.  \label{eq:deLR}
\end{align} 
The detailed analytic formulas of particular contributions listed in Eq.\eqref{eq:deLR} are derived easily based on general results introduced in Ref. \cite{Hue:2024rij}, so we omit here.  

\subsection{$Z\to e_b^\pm e_a^\mp$ decays}

{The effective amplitude for the decay $Z(q) \to e^\pm_a (q_1) e^{\mp}_b (q_2)$ is given in Refs.  \cite{Jurciukonis:2021izn, DeRomeri:2016gum}, and the corresponding partial decay width is presented in detail in our previous work \cite{Hong:2024yhk}}

\begin{figure}[ht]
	\centering 
	\includegraphics[width=16.5cm]{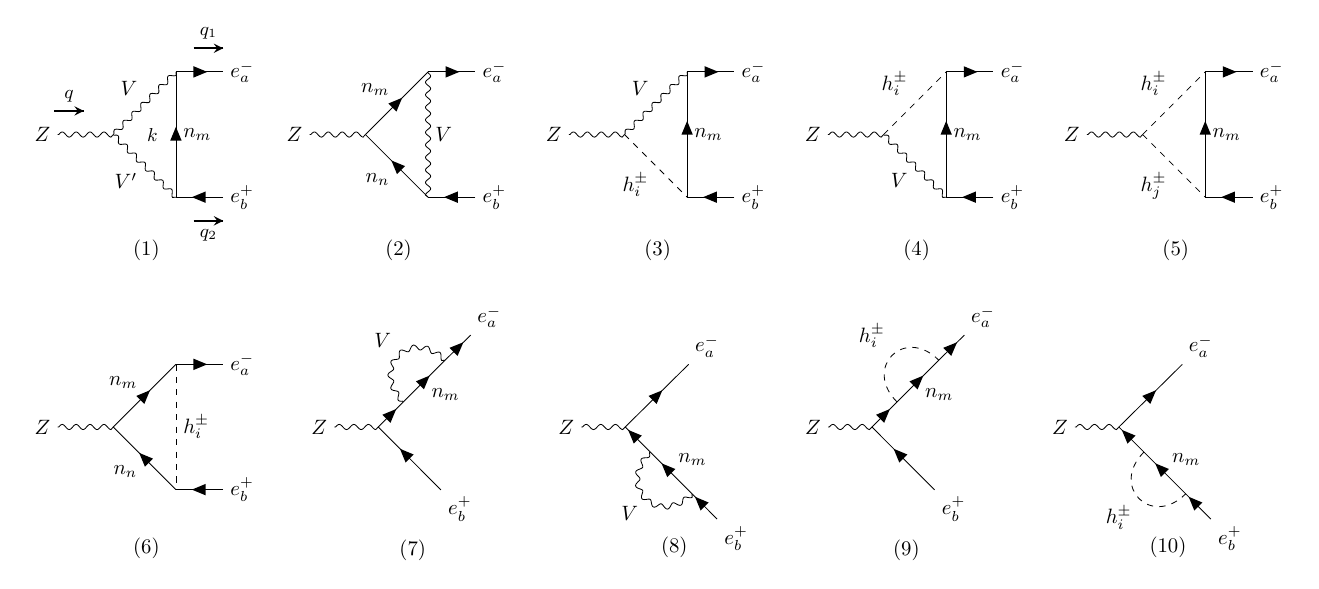}
	\caption{ One-loop Feynman diagrams contributing to {$Z \to e_a^- e_b^+$} in the unitary gauge, where $V=V'=W$.} 
	\label{Zeaeb}
\end{figure}

The relevant one-loop diagrams contributing to the decay amplitude {$Z \to e_a^- e_b^+$} in the unitary gauge are illustrated in Fig. \ref{Zeaeb}.
The {coupling farctors} of gauge bosons and scalars are derived from the Lagrangian covariant kinetic terms, which is showed in Table.~\ref{boson-h}.
\begin{table}[ht]
	\centering 
	\renewcommand{\arraystretch}{1.4}
	\begin{tabular}{cccccccc}
		\hline
		Vertex & factor & Vertex & factor & Vertex & factor &Vertex & factor\\
		\hline
		\hline
		$g_{Z h^+_1h_1^-}$	& {$\dfrac{c_{\alpha _{\pm }}^2-2 s_W^2}{2 c_W s_W}$}	& $g_{Z h^+_2h_2^-}$	& {$\dfrac{s_{\alpha _{\pm }}^2-2 s_W^2}{2 c_W s_W}$}	&$g_{Z h^+_1h_2^-}$	& {$\dfrac{s_{\alpha_{\pm}} c_{\alpha_{\pm}}}{2s_Wc_W} $}&$g_{Z h_i^\pm V^\mp}$ & $0$\\
		\hline			
	\end{tabular}
	\caption{The couplings of gauge bosons and scalars, in which $i=1,2, \, V = W, W_{24}$. \label{boson-h} }
\end{table}

The interactions between the SM gauge bosons $Z$ and leptons derived from Eq. (\ref{eq:lkLep}) can be explicitly expanded as follows 
\begin{align}
	\label{eq:LZll}
	\mathcal{L}_{Z\ell\ell}= & \frac{e}{2s_Wc_W} \overline{e_a}\gamma^{\mu} \left[\left(-1+2s_W^2\right)P_L  +2s_W^2 P_R \right]e_a  Z_{\mu} + \frac{e}{2s_Wc_W} \overline{E_a}\gamma^{\mu} \left[ 2s_W^2  P_L  +2s_W^2 P_R \right]E_a  Z_{\mu} \crn
	&+ \frac{e}{2s_Wc_W} \overline{\nu_{a}}\gamma^{\mu}P_L \nu_{a}  Z_{\mu} +...
\end{align} 

The Lagrangian in the case of the $Z$ with two Majorana leptons $n_m$ and $n_n$ by using {the following  symmetric form \cite{Korner:1992an, Dreiner:2008tw}} 
\begin{align}
	\mathcal{L}_{Z\bar{n}_mn_n} =& \frac{e}{2s_Wc_W} Z_\mu\sum^7_{m,n=1}\bar n_m\gamma^\mu \left[G_{mn}P_L -G_{nm}P_R\right]n_n, 
\;	G_{mn} = G_{nm}^* = \sum^3_{a=1}U_{am}^{\nu*}U_{an}^{\nu}. 
\end{align}

We summarize the interaction between three bosons by using Feynman rules {$-ieg_{ZVV}Z^\mu(q)V^{+\nu}(q_1)V^{-\alpha}(q_2)\Gamma_{\left(\mu\nu\alpha\right)}$} corresponding to the {coupling factors} in Table~\ref{tab_ZVS,SS}, by referring to {Table 1} in Ref.~\cite{Hue:2023VNUHCM}, and we also ignore non-relevant couplings. {Here, $V = W, W_{24}, W_{23}$, and the triple gauge vertex is defined as $\Gamma_{\mu\nu\alpha} (q, q_1, q_2) = g_{\mu\nu}(q-q_1)_\alpha + g_{\nu\alpha}(q_1-q_2)_\mu + g_{\alpha\mu}(q_2-q)_\nu$, where $q, q_1, q_2$ are the corresponding momenta of the gauge bosons.}

\begin{table}[ht]
	\centering 
	\renewcommand{\arraystretch}{1.2}
	\begin{tabular}{cccccc}
		\hline
		Vertex & factor &Vertex & factor &Vertex & factor \\
		\hline
		\hline
		$g_{Z W^+W^-}$	& ${t_W^{-1}} $&  $g_{Z W^+_{24}W^-_{24}}$
		& ${\frac{-1+2s^2_W}{2s_Wc_W}} $ &		$g_{Z W^{0}_{23} W^{0*}_{23}}$
		& ${\frac{-1}{2s_Wc_W}}$\\
		\hline			
	\end{tabular}
	\caption{The coupling factors of the $Z_\mu$ {gauge boson and two charged bosons.}  \label{tab_ZVS,SS} }\setlength{\tabcolsep}{10pt}
\end{table}

The couplings of the photon with charged gauge bosons are determined from the covariant kinetic terms of the non-Abelian gauge fields. The interaction Lagrangian has the general form	
\bea 
-\mathcal{L}^{int}_{kin} = gf^{abc}\partial_\mu A^a_\nu\left(A^{b\mu}A^{c\nu}-A^{c\mu}A^{b\nu}\right),  
\label{LZmuVmu}
\eea
where $f^{abc}$ denote the structure constants. A general formulas to determine all triple and quartic couplings of gauge bosons were given in Ref. \cite{Hue:2023VNUHCM}. Combining the analytic formulas of coupling factors and the general results determining the one-loop contributions to the LFV$Z$ amplitudes given in Ref. \cite{Hue:2024rij}, we easily derive the total one-loop contributions as well as final decay rates Br$(Z\to e^{\pm}_be^{\mp}_a)$.

\subsection{One-loop contributions to $a_{a_e}$ and Br($e_b \to e_a\gamma$)}
\begin{figure}[ht] 
	\includegraphics[width=16.5cm]{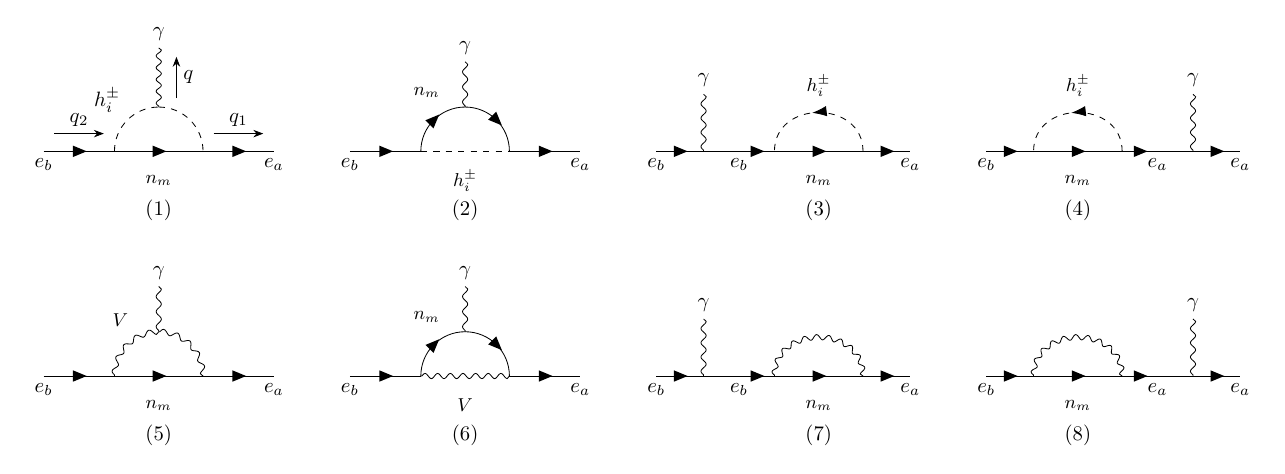}
	\caption{ Feynman diagrams for one-loop contributions to the decay amplitudes$ e_b \to  e_a \gamma$, and $(g-2)_{e_a}$ in the unitary gauge, where $V=W^\pm$. We omit contributions from gauge boson  exchanges $W_{24}$ and $W_{23}$ to $\Delta a_{e_a}$. \label{eba-1LoopRxi}}
\end{figure}

The branching ratios of the cLFV decays are formulated as follows \cite{Lavoura:2003xp, Hue:2017lak, Crivellin:2018qmi}
\begin{align}
\label{eq_brebaga}
\mathrm{Br}(e_b\to e_a\gamma)= \frac{48\pi^2}{G_F^2 m_b^2}\left( \left| c_{(ab)R}\right|^2 + \left| c_{(ba)R}\right|^2\right) \mathrm{Br}(e_b\to e_a \overline{\nu_a}\nu_b),
\end{align}
where $G_F=g^2/(4\sqrt{2}m_W^2)$, Br$(\mu\to e \overline{\nu_e}\nu_{\mu})\simeq 1$,  Br$(\tau\to e \overline{\nu_e} \nu_{\tau}) \simeq 0.1782$, Br$(\tau\to \mu \overline{\nu_\mu}\nu_{\tau})\simeq 0.1739$ \cite{ParticleDataGroup:2020ssz},  and one-loop Feynman diagrams contributing to cLFV decay amplitudes are listed in Fig. \ref{eba-1LoopRxi}. 

In the 341mISS model,  for simplicity we chose $U^N=U^E=I_3$, leading to the following precise expressions of $c_{(ab)R}$: 
\begin{align}
c_{(a b) R}^{341\mathrm{mISS}}= & \sum_{i=1}^2 c_{(a b) R}\left(h^{ \pm}_i\right)+\sum_{V} c_{(a b) R}(V); \; V=W^+,W^+_{24},W^0_{23};
\; c_{(b a) R}=c_{(a b) R}[a \rightarrow b, b \rightarrow a];
 \crn c_{(a b) R}\left(W\right)= &  \sum_{m=1}^7 \frac{g^2 e m_{e_b} U^{\nu}_{am} U^{\nu*}_{bm} \tilde{f}_{V}\left(x_{m, W}\right)}{32 \pi^2 m_W^2}; \; 
 %
 c_{(a b) R}\left(V\right)=   \frac{g^2 em_{e_b}  \delta_{ab} \tilde{f}_{V}\left(x_{f,V}\right)}{32 \pi^2 m_{V}^2}; {f=E_a, N_a};
 \crn c_{(a b) R}\left(h_i^{ \pm}\right)= &  \sum_{m=1}^7 \frac{g^2 e \left[\lambda_{am}^{L, i *} \lambda_{bm}^{R, i} m_{n_m} f_{\Phi}\left(x_{m, i}\right)+\left(m_{e_b} \lambda_{am}^{L, i *} \lambda_{bm}^{L, i}+m_{e_a} \lambda_{am}^{R, i *} \lambda_{bm}^{R, i}\right) \tilde{f}_{\Phi}\left(x_{m, i}\right)\right]}{32 \pi^2 m_W^2 m_{h_i^{ \pm}}^2};
\end{align}
with $x_{m, W} \equiv m_{n_m}^2 / m_{W}^2$, $x_{E_a, W_{24}} \equiv m_{E_a}^2 / m_{W_{24}}^2$, $x_{N_a, W_{23}} \equiv m_{N_a}^2 / m_{W_{23}}^2$,  $x_{m, i} \equiv m_{n_m}^2 / m_{h_i^{ \pm}}^2$. Three  master functions  $\tilde{f}_{V}(x)$, $\tilde{f}_{\Phi}(x)$, and $f_{\Phi}(x)$ and the analytic formulas for these contributions were  shown in Refs. \cite{Thao:2023gvs, Crivellin:2018qmi}. Similarly, the one-loop contributions from the singly charged Higgs boson $h^\pm_{i}$ and $V$ exchanges to $a_{e_a}$ are
\begin{align}
\label{eq_Hpm1}
\Delta a_{e_a}^{341\mathrm{mISS}}&=-\frac{4m_{a}}{e} \mathrm{Re}\left[c_{(aa) R}^{341\mathrm{mISS}} -c_{(aa) R}^{\mathrm{SM}}\right] 
\simeq a^{{h^\pm_i}}_{e_a}=-\frac{4m_{a}}{e} \sum_{i=1}^2 \mathrm{Re}\left[c_{(aa)R}(h^\pm_i)\right],
\end{align}
where $i=1,2; \, a=1,2,3$.

\section{\label{sec_numer} Numerical Investigation}
In this section, we will use the neutrino oscillation data {provided} in Refs. \cite{ParticleDataGroup:2024cfk, Esteban:2024eli}. The lepton mixing matrix $U_{\mathrm{PMNS}}$ has the standard form \cite{ParticleDataGroup:2024cfk} as a function of parameters defined from the experimental data, namely the three mixing angles $\theta_{ij}$ \cite{ParticleDataGroup:2024cfk}, one Dirac phase $\delta$ and two Majorana phases $\alpha_{1}$ and $\alpha_2$~\cite{ParticleDataGroup:2018ovx}, in particular, we assume a normal mass ordering with   $0=m_{n_1}<m_{n_2}<m_{n_3}$, and adopt the following numerical input values:
\begin{align}
	& s_{12}^2=0.307, s_{23}^2=0.561, s_{13}^2=0.0219, \delta=180 \,[\mathrm{Deg}],
\crn & \Delta m_{21}^2=7.49 \times 10^{-5}\left[\mathrm{eV}^2\right], \Delta m_{32}^2=2.459 \times 10^{-3}\left[\mathrm{eV}^2\right],
\crn 	\hat{m}_\nu & =\left(\hat{m}_\nu^2\right)^{1 / 2}=\operatorname{diag}\left(0, \sqrt{\Delta m_{21}^2}, \sqrt{\Delta m_{21}^2+\Delta m_{32}^2}\right), \\
U_{\mathrm{PMNS}} & =\left(\begin{array}{ccc}
c_{12} c_{13} & s_{12} c_{13}  & s_{13} e^{-i \delta_{CP}} \\
-s_{12}c_{23} -c_{12} s_{13} s_{23} e^{i \delta_{CP}} & c_{12} c_{23}-s_{12} s_{13} s_{23} e^{i \delta_{CP}} & c_{13} s_{23} \\
s_{12} s_{23}-c_{12}s_{13} c_{23}  e^{i \delta_{CP}} & -{s_{12} s_{13}c_{23}  e^{i \delta_{CP} }}-c_{12} s_{23} & c_{13} c_{23}
\end{array}\right).
\end{align}
The experimentally determined parameters are \cite{ParticleDataGroup:2024cfk}: $g  =0.652, G_F=1.166378 \times 10^{-5} \mathrm{GeV}, s_W^2=0.231, m_W=80.3692 \mathrm{GeV}$, $	m_e  =5 \times 10^{-4} \mathrm{GeV}, m_\mu=0.105 \mathrm{GeV}, m_\tau=1.777 \mathrm{GeV}, m_Z=91.1880 \mathrm{GeV}$. 
The scanning ranges of free parameters are chosen as follows:
\begin{align}
\label{eq:scanning}
	&m_{n_{4,5}}  \in[0.1,5] \mathrm{TeV}, m_{h_{1,2}^{ \pm}} \in[0.5,5] \mathrm{TeV},
\;  t_\beta  \in[5,25], \; x_0 \in\left[10^{-6}, 5\times  10^{-4}\right], 
\crn &
-\pi \leq \alpha_{\pm},\; \zeta \leq \pi,\;\delta =0,\;  |y^s_{ia}|\leq 3 \forall i=1,2, \;a=1,2,3,
\end{align}
where the Yukawa couplings $y^s_{ia}$ alway satisfy the perturbative limit, and $x_0$ is defined in the  NO scheme as: 
 	\begin{equation}\label{eq:x0}
 	\hat{x}_\nu\equiv \frac{\hat{m}_\nu}{\mu_0}=x_0\times \mathrm{diag}\left( 0, \frac{m_{n_2}}{m_{n_3}},1\right) , \; x_0\equiv  \frac{m_{n_3}}{\mu_0}.
 \end{equation}
The upper bound of $x_0$ is constrained from the non-mixing property of the part $\left(\mathbb{I}_3-\frac{1}{2}\hat{x}_\nu\right)$ appearing in the $U^{\nu}$ given in Eq. \eqref{eq_Unu0} \cite{Fernandez-Martinez:2016lgt, Agostinho:2017wfs, Blennow:2023mqx}.
We note that the singly charged Higgs bosons have masses at the TeV scale because they are associated with the $SU(3)_L$ and $SU(4)_L$ symmetry-breaking scales, as well as the parameter $\mu_5^2$ in the Higgs potential \eqref{eq_Vh}, as can be seen from the squared mass matrix in Eq.  \eqref{eq:mc2}. The Higgs self-coupling contributing  to the coupling factors $\lambda_{h_{ij}}$ appearing in the diagram (5) of Fig. \ref{fig_LFVh} has been verified to satisfy the theoretical constraints from perturbative unitarity and vacuum stability conditions  in the Higgs potential. This conclusion is supported by recent studies on 3-3-1 models, where analogous  Higgs self-couplings play the same role as in the 341mISS model \cite{Sanchez-Vega:2018qje, Costantini:2020xrn, Kannike:2025qru}.
 	
Regarding our numerical investigation, all collected parameter points  satisfy the experimental bounds listed in Table \ref{dataLFV}, the $1\sigma$ deviation from the SM prediction for $\Delta a_{\mu} \equiv \Delta a_{\mu}^{341\mathrm{mISS}}$ given in Eq. \eqref{eq:damu}, as well as the  constraint  $10^{-14}\leq |\Delta a_e|\equiv |\Delta a_{e}^{341\mathrm{mISS}}|\leq 8\times 10^{-13}$. We specifically focus on the regions of the parameter space that yield large $|\Delta a_{\mu}|\geq 10^{-10}$. Regions with smaller value  $|\Delta a_{\mu}|< 10^{-10}$ exhibit LFV properties similarly to  those with  $|\Delta a_{\mu}|\simeq  10^{-10}$, and these points are readily obtained; consequently, they are not presented here.

 To obtain the allowed parameter points, we adopt the following scanning procedure. In the first step,  all independent parameters  are generated randomly within the ranges  specified in Eq. \eqref{eq:scanning}. This yields  an  initial parameter set, $\{ z^0_1=t_{\beta},\; z^0_2=m_{n_4},\dots\}$, satisfying the preliminary conditions: $|\Delta a_{\mu}|\geq 5\times 10^{-11}$, Br$(\mu\to e\gamma)<10^{-8}$, Br$(\tau\to \mu \gamma, e\gamma)<10^{-6}$, Br$(Z\to \mu e)<10^{-4}$, Br$(h\to \mu e)<10^{-3}$. In addition, all remaining bounds listed in Table \ref{dataLFV}, together with the theoretical constraints on the model parameters and couplings,  including perturbative limits, perturbative unitarity, and vacuum stability, are required to be satisfied. In the second step, a new set of parameters is generated randomly in the vicinity of the initial point. For example  $t_{\beta}=z_1\in [ z_1^0-0.1 |z_1^0|,z_1^0+0.1 |z_1^0|],\; \dots$, and similarly for the remaining parameters. The purpose of this local scan is to identify parameter sets yielding larger values of $|\Delta a_{\mu}|$, while simultaneously reducing the LFV decay rates  Br$(e_b\to e_a\gamma)$, and Br$(Z,h\to \mu e)$. This procedure is repeated until a parameter set satisfying $|\Delta a_{\mu}|\geq 10^{-10}$, together with all experimental bounds on $(g-2)_{e_a}$ and LFV decays, is found, or until the number of sampled points becomes excessively large. The algorithm then returns to the first step to search for additional allowed points. Our numerical analysis is based on approximately 120 allowed parameter points obtained after scanning several tens of millions of points within parameter ranges specified in Eq. \eqref{eq:scanning}.

\begin{figure}[ht]
		\centering
		\begin{tabular}{c}
\includegraphics[width=8.2cm]{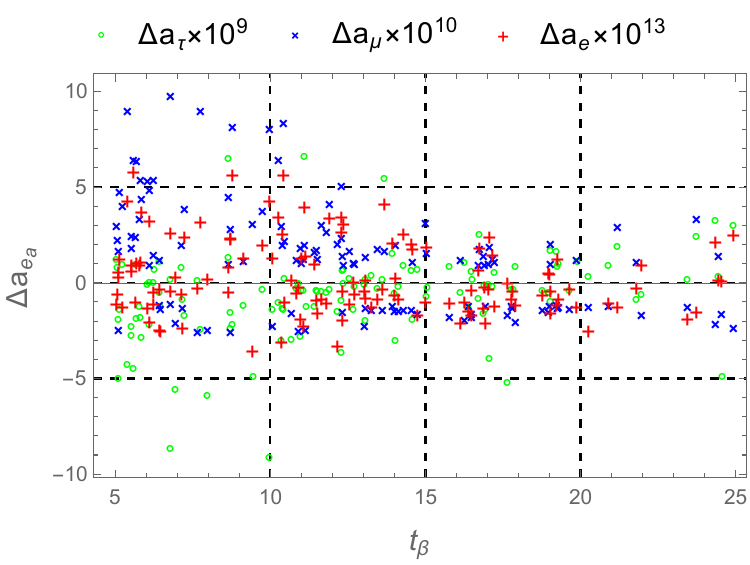}
	\includegraphics[width=8.5cm]{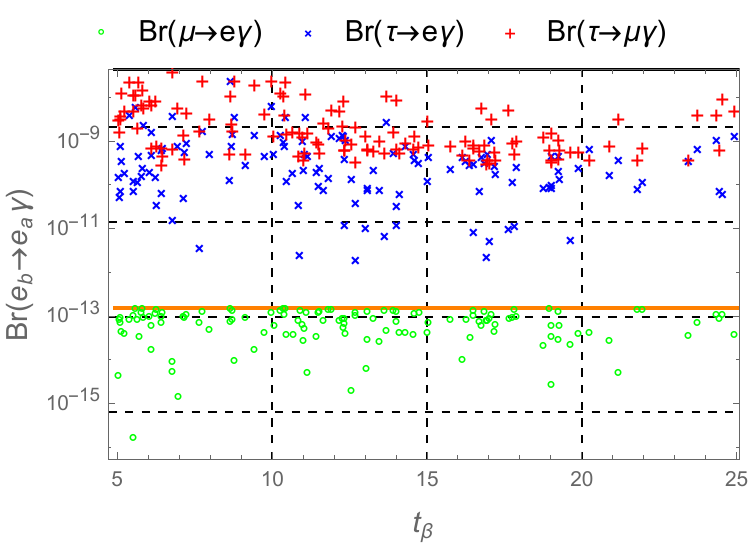} \\
\includegraphics[width=8.cm]{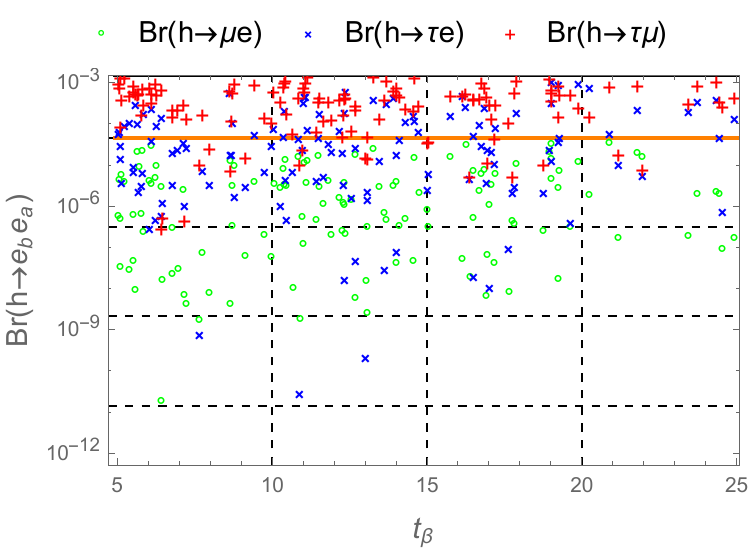}
\includegraphics[width=8.cm]{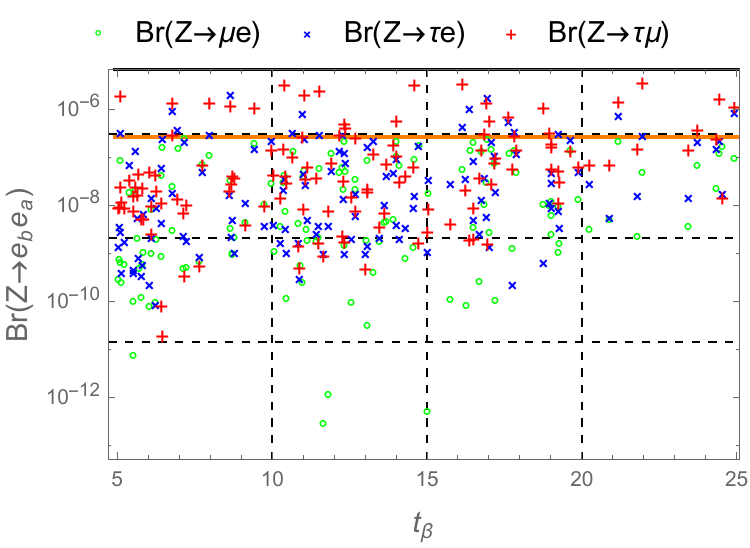} \\
		\end{tabular}
		\caption{The dependences of  $\Delta a_{e_a}$ and LFV decay rates  with  $t_\beta$. In each panel, the two black and orange horizontal lines denote the current experimental upper bounds listed in Table~\ref{dataLFV}. The corresponding values are: $4.2\times 10^{-8}$ and $1.5\times 10^{-13}$ for $\text{Br}\left(e_b\to e_a\gamma\right)$, respectively; $2.0\times 10^{-3}$ and $4.4\times 10^{-5}$ for $\text{Br}\left(h\to e_be_a\right)$, respectively; and $6.5\times 10^{-6}$ and $2.62\times 10^{-7}$ for $\text{Br}\left(Z\to e_be_a\right)$, respectively. }\label{fig:tbLFV}
	\end{figure}

We begin the presentation of our numerical results by showing the  dependence of $\Delta a_{e_a}$ and  LFV decay rates on  $t_{\beta}$ in Fig. \ref{fig:tbLFV}.
In general, all allowed values for $\Delta a_{e_a}$ and LFV decay rates remain consistent with both current experimental bounds  and projected future sensitivities. Notably, regions with large $t_{\beta}$ tend to increase the lower bounds of LFV decay rates. The maximal value $|\Delta a_{\tau}|\leq {\mathcal{O}(10^{-8})}$ is still well below  the recent experimental sensitivity {\cite{Bernabeu:2007rr, Chen:2018cxt, Denizli:2024uwv}}.  The most interesting property we realize here is that large values close experimental constraints  of both $\Delta a_{\mu}$ and Br$(\tau \to \mu \gamma)$ prefer small $t_{\beta}$, suggesting a strong correlation between these two physical quantities. On the other hand, the Br$(\mu\to e\gamma)$ does not impose significant constraints on the  remaining quantities, despite being the most stringently restricted channel in both current and forthcoming experimental searches.

\begin{figure}[h]
	\centering
	\begin{tabular}{c}
		\includegraphics[width=8.2cm]{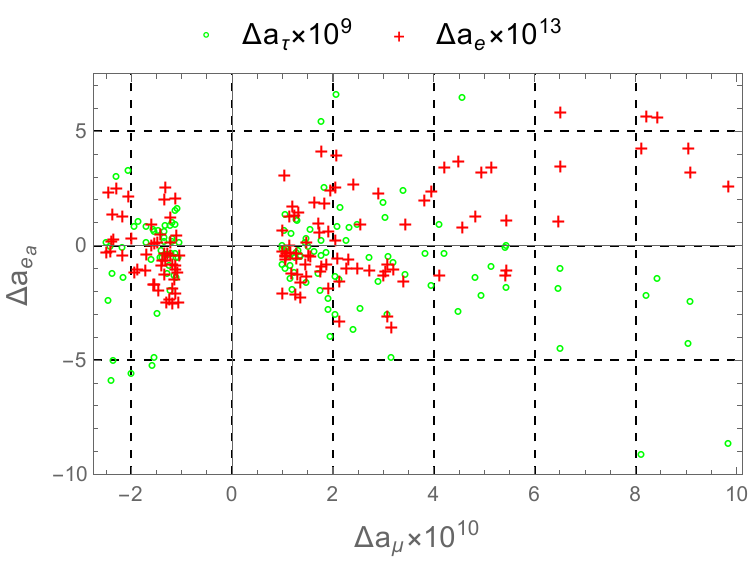}
		\includegraphics[width=8.5cm]{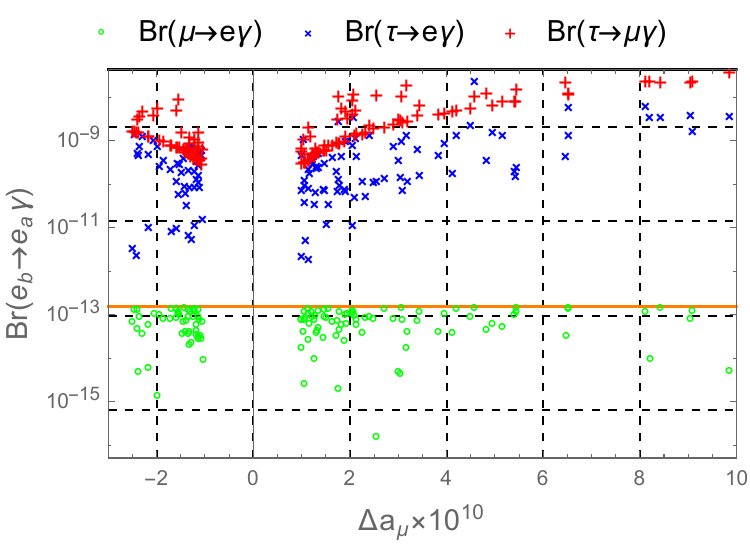}
		\\
		\includegraphics[width=8.5cm]{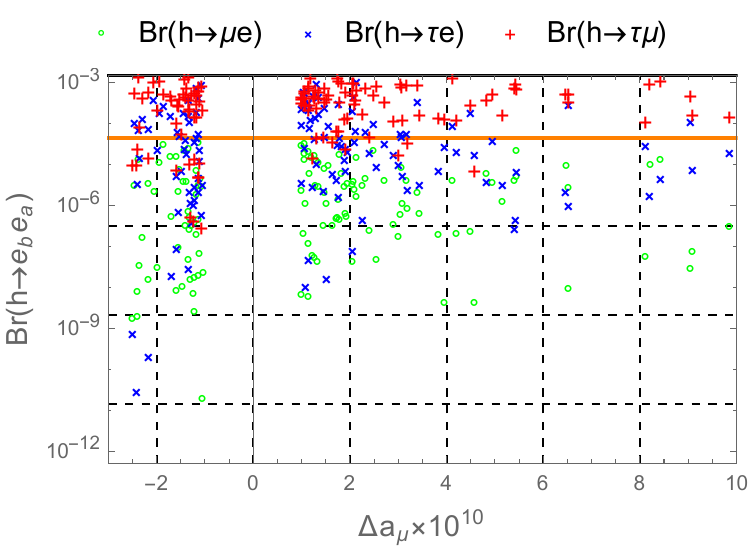}
		\includegraphics[width=8.5cm]{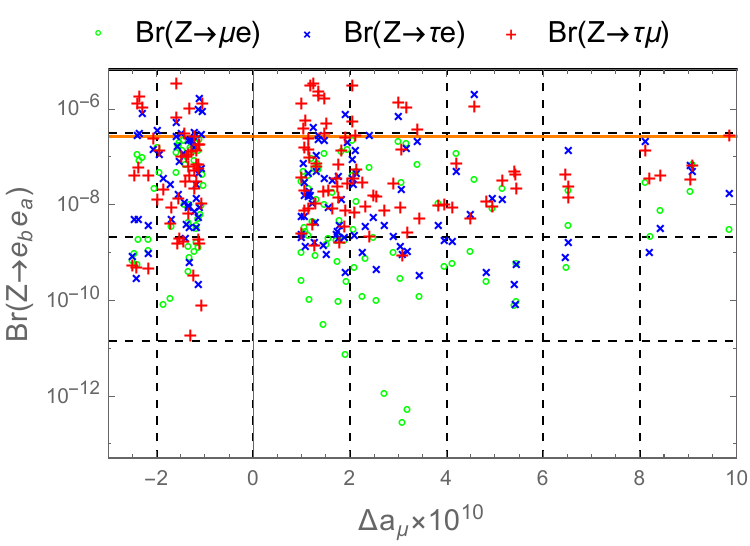}
		 \\
	\end{tabular}
	\caption{The relations between   $\Delta a_{e,\tau}$ and LFV decay rates  with  $\Delta a_{\mu}$. In each panel, the two black and orange horizontal lines denote the current experimental upper bounds listed in Table~\ref{dataLFV}. The corresponding values are: $4.2\times 10^{-8}$ and $1.5\times 10^{-13}$ for $\text{Br}\left(e_b\to e_a\gamma\right)$; $2.0\times 10^{-3}$ and $4.4\times 10^{-5}$ for $\text{Br}\left(h\to e_be_a\right)$; and $6.5\times 10^{-6}$ and $2.62\times 10^{-7}$ for $\text{Br}\left(Z\to e_be_a\right)$, respectively. }\label{fig:am_aetLFV}
\end{figure}

Fig. \ref{fig:am_aetLFV} focuses on the relations of $\Delta a_{e,\tau}$ and LFV rates with $\Delta a_{\mu}$.
Several compelling features can be identified from our analysis. First, small  $|\Delta a_{\mu}|$ around $10^{-10}$ values  predict wide allowed ranges of all mentioned quantities, as is typical in many BSM scenarios. In contrast, as $\Delta a_\mu$ approaches the $1 \sigma$ upper limit of $10^{-9}$, stringent constraints emerge; notably, $\text{Br}(\tau \to \mu \gamma)$ saturates the current experimental upper bound.   The forthcoming sensitivity of Br$(\tau \to \mu \gamma)$ given in Table \ref{dataLFV} will result in the largest value $\Delta a_{\mu}\leq 5\times 10^{-10}$. This indicates that any improvement in the experimental precision of the $\tau \to \mu \gamma$ decay will lead to substantially stronger constraints on $(g-2)_{\mu}$ data.  The  remaining two channels show a much weaker dependence on the muon anomalous magnetic moment (AMM), but only $3.5\times 10^{-16}\leq\text{Br}(\mu\to e\gamma)\leq 1.5\times 10^{-13}$ is suitable with current data value and near future sensitivity. In the right panel, only $\text{Br}(\mu\to e\gamma)$ reaches the current experimental upper bound, while the other two  decay rates  remain below this limit.

Regarding the dependence of LFV$h$ and LFV$Z$ decays on $\Delta{a_\mu}$, as shown in the two lower panels of Fig.~\ref{fig:am_aetLFV}. In the first panel, for $\Delta a_\mu \geq 1.0 \times 10^{-10}$, both {$\text{Br}(h\to \mu e)$ and $\text{Br}(h \to \tau \mu)$ approach the experimental limits data in two solid lines, respectively. In addition, only the $\text{Br}(h\to \tau e) \leq 8\times 10^{-4}$ are remains below than with $2.0\times 10^{-3}$ of current value. In the second panel, a similar behavior is observed: LFV$Z$ decay rates are not strongly tightly bound with $\Delta a_\mu$. Furthermore, only $\text{Br}(Z \to \mu e)$ reaches the latest experimental sensitivity, whereas the other Br remain below their current experimental limits.

Fig.~\ref{fig_LFVae} illustrates the relationships between the LFV decay rates and $\Delta a_e$.
\begin{figure}[ht]
	\centering
	\includegraphics[width=9cm]{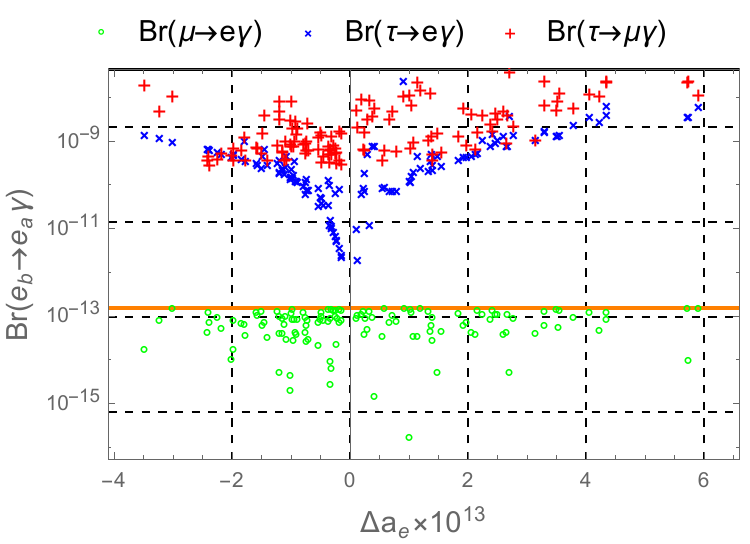}
\begin{tabular}{cc}
	\includegraphics[width=8.5cm]{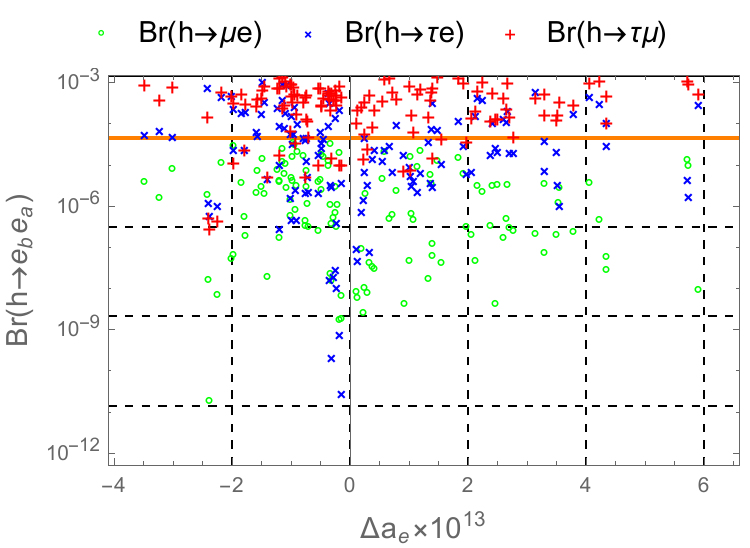} &
	\includegraphics[width=8.5cm]{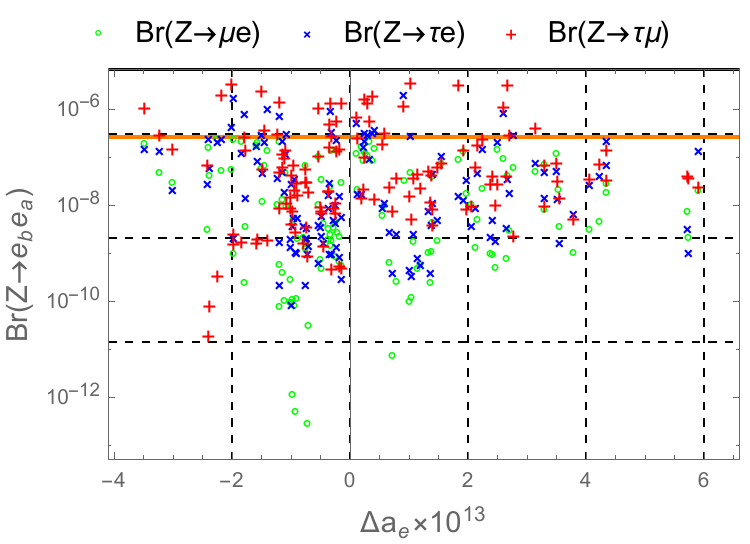} 
\end{tabular}
	\caption{The relationship between decay rates of cLFV, LFV$h$, and  LFV$Z$ with respect to $\Delta{a_e}$. In each panel, the two black and orange horizontal lines denote the current experimental upper bounds listed in Table~\ref{dataLFV}. The corresponding values are:  $4.2\times 10^{-8}$ and $1.5\times 10^{-13}$ for $\text{Br}\left(e_b\to e_a\gamma\right)$; $2.0\times 10^{-3}$ and $4.4\times 10^{-5}$ for $\text{Br}\left(h\to e_be_a\right)$; and $6.5\times 10^{-6}$ and $2.62\times 10^{-7}$ for $\text{Br}\left(Z\to e_be_a\right)$, respectively.}\label{fig_LFVae}
\end{figure}
 The results show that {both three channels of the LFV$h$ decays} do not change much when $\Delta a_e$ increases, but $\text{Br}(h\to \tau\mu), \, \text{Br}(h\to \mu e)$ still can get the near latest data values of them. Though the last channel decay in LFV$h$ is lower than the current data, it will reach the coming sensitivity.  Similarly, all  LFV$Z$ decay rates have little change with  increasing $\Delta a_e$  in the region $\in[2.4\times 10^{-14},\, 1.1\times 10^{-13}]$. However, $\text{Br}(Z\to \tau e)$ is still quite far from the limit bound by the latest data value, with only one of these channels reaching the upper bound of the new updated experiment.

Finally, we  focus on the correlations between the  cLFV decay rates  and $\Delta a_e$ presented in the first panel of  Fig.~\ref{fig_LFVae}. Although $\text{Br}(\mu\to e\gamma)$  depends weakly on $\Delta a_e$, it can reach the current experimental upper bound $1.5\times 10^{-13}$, and satisfies the  future sensitivity of  $4.0\times 10^{-16}$ \cite{MEGII:2025gzr,MEGII:2018kmf,Belle-II:2018jsg}. Similarly to the  case of $\text{Br}(\tau\to \mu\gamma)$. On the other hand, although $\text{Br}(\tau \to e \gamma)$ has not yet reached the current experimental bound, it exhibits a strong correlation with $\Delta a_e$. Consequently, this decay channel stands as a promising candidate to be observed in the near future as experimental sensitivities continue to refine.

Besides the radiative LFV decays considered in this work, the present framework also induces other LFV observables such as $\mu \to 3e$, $\tau \to 3\ell$, and $\mu - e$ conversion in nuclei;     see, for example, the comprehensive review in Ref. \cite{Lindner:2016bgg}, which also discusses 3-3-1 models with ISS neutrinos. The current experimental bounds on these LFV  processes are still at the levels of  $\mathcal{O}(10^{-8})$ and  $\mathcal{O}(10^{-12})$ for the corresponding $\tau$- and $\mu$-related channels \cite{Belle-II:2024sce, Belle-II:2025urb}. These bounds are generally less  stringent than those from the cLFV processes; see the summary in Ref. \cite{COMET:2025sdw}. In the 341mISS model, LFV processes do not have tree-level contributions. Consequently, the leading contributions arise at the  one-loop level, where the corresponding amplitudes can be expressed in terms of linear combinations of one-loop three-and four-point Passarino-Veltman functions, multiplied by products of four vertex-factors containing the LFV sources. Therefore, one may expect the resulting predictions for the three-bodyLFV decay  and $\mu-e$ conversion rates to be  smaller than those for the cLFV rates considered in this work.   Consequently,  current experimental constraints on three-body LFV decays and $\mu-e$ conversion do not exclude the allowed regions of parameter space we identified in our analysis. A complete analytical and numerical investigation of these processes, combined with the LFV observables studied here, is required to determine more stringent constraints on the parameter space and will be presented in future work.

\section{\label{sec_con} Conclusions}

We have investigated the 341mISS model, which extends the original 3-4-1 model by introducing a new singly charged Higgs boson and mISS neutrinos, thereby providing an explanation for the large values of $(g-2)_{e,\mu}$ that remain compatible with recent experimental results. Unlike the 341ISS model previously studied in Ref. \cite{Thao:2023gvs}, the present work considers a 341mISS realization incorporating only four heavy right-handed neutrinos. We demonstrate that this minimal version is sufficient to generate sizable one-loop contributions that accommodate large $|\Delta a_{e,\mu}|$ and LFV decay rates consistent with experimental constraints.
 
First, we identify the allowed regions of  the parameter space where $\Delta a_\mu$ can reach $10^{-9}$, aligning with the latest (2025) experimental data within a $1\sigma$ deviation from the SM prediction. Simultaneously, enhanced values of $|\Delta a_e| \sim \mathcal{O}(10^{-13})$ are attainable. Second, we  demonstrate the correlations between LFV decay rates and the $\Delta {a_{e,\mu}}$ {constraints}. Our results reveal a significant dependence of cLFV ratios on these {constraints}, which helps tighten the upper bounds for  these observables in light  of future experimental sensitivities. 

 In summary, several LFV decay rates among the cFLV, LFV$h$, LFV$Z$ decays are in good agreement with current and forthcoming experimental data, including $\mu \to e\gamma, \, \tau \to \mu\gamma, \, h\to \mu e, \, Z\to \mu e, \, Z \to \tau\mu $. Conversely, $\text{Br}(\tau \to e\gamma)$ remains significantly below forthcoming experimental sensitivity. Our numerical analysis indicates that any future updated information of the $\tau \to \mu \gamma$ decay would impose more stringent constraints on $\Delta a_{\mu}$ than the current $1\sigma$ experimental deviation. Notably, this strong correlation renders the 341mISS model significantly more predictive than previous studied versions. The viability of the model can be further tested as updated experimental data on  $(g-2)_{\mu,e}$ AMM and cLFV decay rate $\text{Br}(\tau \to \mu \gamma)$ become available.

\section*{Acknowledgments}
{The authors are grateful to Dr. Bibhabasu De, Dr. Andrea Sainaghi, Dr. Pengxuan Zhu, Prof. Arindam Das, and Prof. Maxim KHLOPOV for helpful discussions.} This research is funded by Vietnam National
University HoChiMinh City (VNU-HCM) under grant number B2026-16-02.
\appendix
\section{\label{app_gaugebosons} Gauge boson masses and mixing parameters}
The relation between the original basis $(W_3, W_8,W_{15}, B'')$ and the mass basis $(A,Z,Z_3,Z_4)$  of all real neutral boson was determined previously \cite{Long:2016lmj}. 
 The  masses of non-Hermitian  gauge bosons are given by
 \begin{align}
 m^2_W &  =  \fr{g^2(v_1^2 + v_2^2)}{4}\, ,  m^2_{W_{13}}  =  \fr{g^2(v_2^2 + v_\om^2)}{4}\, ,
 m^2_{W_{23}}  =  \fr{g^2(v_1^2 + v_\om^2)}{4}\, , \crn
 m^2_{W_{14}} & =     \fr{g^2(v_2^2 + v_\chi^2)}{4}\, , m^2_{W_{24}}  =  \fr{g^2(v_1^2 + v_\chi^2)}{4}\, ,
 m^2_{W_{34}}  =  \fr{ g^2(v_\om^2 + v_\chi^2)}{4}\, .
 \label{eq_mWpm}	
 \end{align}
 By  spontaneous symmetry breaking (SSB), the following relation should be in order:
 $v_\chi\gg v_\om \gg v_1, v_2$.  A consequence from Eq. \eqref{eq_mWpm} is  that $W^\pm$ must be identified with the singly charged  SM gauge boson, namely 
 \be v_1^2 + v_2^2 = v^2= 246^2\; \textrm{GeV}^2. \ee

 The above formula implies that $\eta$ and $\rho$ play roles of two Higgs doublets in the well-known two Higgs doublet models  after the breaking steps to the SM gauge group  $SU(2)_L\times U(1)_Y$.  Then we define the mixing angle $\beta$  as follows
 \begin{equation}\label{eq_tb}
 t_{\beta}\equiv \tan\beta= \frac{v_2}{v_1},\; v_1=v c_{\beta}, \; v_2=vs_{\beta},
 \end{equation}
 where $t_{\beta}\geq0.4$ and  $t_{\beta}<390$ from the contraints of the  Yukawa couplings of  top quark and   tau mass, respectively.
 
Investigating the neutral CP-odd Higgs bosons, we focus on a $4\times4$ matrix for their masses in the imaginary part of base $\left(\rho_2^0, \phi_3^0, \eta_1^0, \chi_4^0 \right)$. It has one zero value corresponding to the photon $A_\mu$ field, and another has non-zero values corresponding to $Z_\mu, Z_{3\mu}, Z_{4\mu}$ neutral gauge bosons. Namely, the neutral gauge boson masses are
 \begin{equation}
m_Z^2 \simeq \frac{m_W^2}{c_W^2}, m_{Z_3}^2 \simeq \frac{4 g^2 c_W^2 v^2_\om}{3\left(1-4 s_W^2\right)}, m_{Z_4}^2 \simeq \frac{g^2\left(9 v^2_\chi+v^2_\om\right)}{6}.
 \label{eq_mV0}	
 \end{equation}
where $s_W^2 = \text{sin}^2\theta_W \simeq 0.231$, with $\theta_W$ is Weinberg angle.

The relations between the flavor and physical base of the neutral gauge bosons is \cite{Long:2016lmj}, but we focus on a simple scenario with $ t_{2\theta}  = \frac{s_{2\theta}}{c_{2\theta}} \propto \mathcal{O}(v_\om^2/v_\chi^2) \simeq 0$. As a result, we collected the following relationship
 \begin{equation}\label{eq_NG0}
 	\begin{pmatrix}
 		W_{3\mu}\\ 
 		W_{8\mu}\\ 
 		W_{15\mu}\\ 
 		B''_{\mu}
 	\end{pmatrix} =
 	\begin{pmatrix}
 		 s_W A_{\mu }+c_W Z_{\mu }\\
 		 c_W c_{32} A_{\mu } -s_W c_{32} Z_{\mu }-s_{32} Z_{3\mu }\\
 		 c_W s_{32}c_{43} A_{\mu } -s_W s_{32}c_{43} Z_{\mu} +c_{32} c_{43} Z_{3\mu } -s_{43} Z_{4\mu }\\
 		 c_W s_{32} s_{43}A_{\mu } -s_Ws_{32} s_{43} Z_{\mu} +c_{32} s_{43} Z_{3\mu } +c_{43} Z_{4\mu } 
 	\end{pmatrix}, 
 \end{equation}
 where 
 \begin{align}
 	\label{eq_sijNG}
 	s_{43}&= \sqrt{\frac{3- 6s^2_W}{3 -4 s_W^2}} , \; c_{43}=-\sqrt{1-s_{43}^2},
 	s_{32} = \sqrt{\frac{3- 4s^2_W}{3c^2_W}}, \;  c_{32}= \sqrt{1-s_{32}^2},
 \end{align} 
 \section{\label{potential}Higgs potential and Higgs spectrum}
The Higgs potential in this work includes two part that four first four lines for the 341 original model and the last two lines for the new singly charged Higgs boson $s^\pm$, namely
 \begin{align} 
 \label{eq_Vh}
V_h =& \mu^2_1 \eta^\dag \eta +
 \mu^2_2  \rho^\dag \rho +  \mu^2_3  \ph^\dag \ph + \mu^2_4  \chi^\dag \chi  
 +\lambda_1 (\eta^\dag \eta)^2 + \lambda_2(\rho^\dag \rho)^2 +
 \lambda_3  (\ph^\dag \ph)^2  + \la_4 ( \chi^\dag \chi)^2 \crn
 & +  (\eta^\dag \eta) [ \lambda_5 (\rho^\dag \rho) +
 \lambda_6 (\ph^\dag \ph)  + \lambda_7 (\chi^\dag \chi) ] + \ (\rho^\dag \rho)[ \lambda_8(\ph^\dag \ph) + \la_9 (\chi^\dag \chi)] +\lambda'_9 (\ph^\dag \ph) (\chi^\dag \chi) \crn
 & +  \lambda_{10} (\rho^\dag \eta)(\eta^\dag \rho) +
 \lambda_{11} (\rho^\dag \ph)(\ph^\dag \rho) + \lambda_{12} (\rho^\dag \chi)(\chi^\dag \rho)   +   \lambda_{13} (\ph^\dag \eta)( \eta^\dag \ph) +\la_{14}\, (\chi^\dag \eta)( \eta^\dag \chi)
\crn
 & +\la_{15}\, (\chi^\dag \ph)( \ph^\dag \chi)  + (f \ep^{i j k l} \eta_i \rho_j \ph_k \chi_l +{\mathrm{h.c.}}) +\mu_5^2 s^+ s^-+\lambda_s\left(s^+ s^-\right)^2 +\left(f_s s^+ \rho^{\dagger} \eta+{\mathrm{h.c.}}\right) \crn
	&+\left(s^+ s^-\right)\left(\lambda_{s \eta} \eta^{\dagger} \eta+\lambda_{s \rho} \rho^{\dagger} \rho+\lambda_{s \phi} \phi^{\dagger} \phi+\lambda_{s \chi} \chi^{\dagger} \chi\right) .
 \end{align} 
 The detailed discussion to derive the masses and mixing parameters of the Higgs bosons was previously presented in Ref. \cite{Long:2016lmj}. Therefore, in this work, we provide brief summary of the masses and mixing of relevant charged scalars. 
 
The mixing of $\chi_2^{\pm}$ and $\rho^\pm_4$ results in a massless Goldstone boson $G^{\pm}_{24}$ of  $W^{\pm}_{24}$ and  a singly charged Higgs bosons $h^{\pm}_3$:
 \begin{align}
 \label{eq_chi2rho4}
 \begin{pmatrix}
 \chi_2^{\pm}	\\
 \rho_4^{\pm}	
 \end{pmatrix}
 =\left(
 \begin{array}{cc}
 c_{\theta_1} & s_{\theta_1} \\
 -s_{\theta_1} & c_{\theta_1} \\
 \end{array}
 \right)	\begin{pmatrix}
 G_{24}^{\pm}	\\
 h_3^{\pm}	
 \end{pmatrix},\;  M^2_{h^\pm_3}= \left(v^2_1 +v_\chi^2\right) \left(\frac{\lambda_{12}}{2}-\frac{f t_{\beta } v_\om}{2 v_\chi}\right), 
 \end{align}
 
 and 
 \begin{equation}\label{eq_theta1}
 s_{\theta_1}\equiv \sin\theta_1,\; 	c_{\theta_1}\equiv \cos\theta_1,\;  t_{\theta_1}\equiv \tan\theta_1=\frac{v_1}{v_\chi}=\frac{vc_{\beta}}{v_\chi}. 
 \end{equation}

We consider  here the mixing of $\left(\phi_1^{\pm}, \eta^\pm_3\right)$, which leads to a Goldstone boson $G^{\pm}_{13}$  of  $W^{\pm}_{13}$ and  a singly charged Higgs bosons $h^{\pm}_{13}$:
 \begin{align}
 \label{eq_chi2rho4}
 \begin{pmatrix}
 \phi_1^{\pm}	\\
 \eta_3^{\pm}	
 \end{pmatrix}
 =\left(
 \begin{array}{cc}
 c_{\theta_2} & s_{\theta_2} \\
 -s_{\theta_2} & c_{\theta_2} \\
 \end{array}
 \right)	\begin{pmatrix}
 G_{13}^{\pm}	\\
 h_{13}^{\pm}	
 \end{pmatrix},\;  M^2_{h^\pm_{13}}= \left(v_2^2+v_\om^2\right) \left(\frac{\lambda_{13}}{2}-\frac{ft_{\beta }^{-1} v_\chi}{2 v_\om}\right), 
 \end{align}
  and 
 \begin{equation}\label{eq_theta2}
 s_{\theta_2}\equiv \sin\theta_2,\; 	c_{\theta_2}\equiv \cos\theta_2,\;  \tan\theta_2=\frac{v_2}{v_\om}=\frac{vs_{\beta}}{v_\om}. 
 \end{equation}
The mixing of $\chi_3^{\pm}$ and $\phi_4^\pm$ results in a massless Goldstone boson $G^{\pm}_{34}$  of  $W^{\pm}_{34}$ and  a singly charged Higgs bosons $h^{\pm}_{34}$:
 \begin{align}
 \label{eq_chi2rho4}
 \begin{pmatrix}
 \chi_3^{\pm}	\\
 \phi_4^{\pm}	
 \end{pmatrix}
 =\left(
 \begin{array}{cc}
 c_{\theta_3} & s_{\theta_3} \\
 -s_{\theta_3} & c_{\theta_3} \\
 \end{array}
 \right)	\begin{pmatrix}
 G_{34}^{\pm}	\\
 h_{34}^{\pm}	
 \end{pmatrix},\;  M^2_{h^\pm_{34}}= \left(v_\om^2+v_\chi^2\right) \left(\frac{\lambda_{15}}{2}-\frac{f v^2s_\beta c_\beta}{2 v_\om v_\chi}\right), 
 \end{align}
  and 
 \begin{equation}\label{eq_theta2}
 s_{\theta_3}\equiv \sin\theta_3,\; 	c_{\theta_3}\equiv \cos\theta_3,\;  \tan\theta_3=\frac{v_\om}{v_\chi}. 
 \end{equation}

We just focus on the basis $(\rho^{\pm},\; \eta^{\pm},\; s^{\pm})$, which was not calculated previously 
 \begin{align}
 \label{eq:mc2}
 \left(
 \begin{array}{ccc}
 \frac{s_{\beta } \left(c_{\beta } s_{\beta } \lambda_{10} v^2-f v_{\omega}v_{v_\chi}\right)}{2 c_{\beta }} & \frac{1}{2} \left(c_{\beta } s_{\beta } \lambda_{10} v^2-f v_{\omega}v_{v_\chi}\right) & \frac{f_s s_{\beta }
 	v}{\sqrt{2}} \\
 \frac{1}{2} \left(c_{\beta } s_{\beta } \lambda_{10} v^2-f v_{\omega}v_{v_\chi}\right) & \frac{c_{\beta } \left(c_{\beta } s_{\beta } \lambda_{10} v^2-f v_{\omega}v_{v_\chi}\right)}{2 s_{\beta }} & \frac{c_{\beta } f_s
 	v}{\sqrt{2}} \\
 \frac{f_s s_{\beta } v}{\sqrt{2}} & \frac{c_{\beta } f_s v}{\sqrt{2}} & \frac{1}{2} \left(s_{\beta }^2 \lambda_{s \eta} v^2+c_{\beta }^2 \lambda_{s \rho}
 v^2+\lambda_{s \chi} v_\chi^2+\lambda_{s \phi} v_\omega^2+2 \mu_5^2\right) \\
 \end{array}
 \right)
 \end{align}

Besides, three  singly charged Higgs bosons $( \rho^\pm_1,\, \eta^\pm_2,\, s^\pm)$ are  changed into  the  physical states $h^\pm_{1,2}$ and the Goldstone bosons $G^\pm_W$ of $W^\pm$ as follows:
 \begin{align}
 \label{eq_rho1eta2}
 \begin{pmatrix}
 \rho_1^\pm	\\
 \eta_2^\pm \\
 s^\pm
 \end{pmatrix}
 =\left(
 \begin{array}{ccc}
 c_{\beta} & s_{\beta}c_{\alpha_\pm} & s_{\beta}s_{\alpha_\pm} \\
 -s_{\beta} & c_{\beta}c_{\alpha_\pm} & c_\beta s_{\alpha_\pm} \\
  0 & -s_{\alpha_\pm} & c_{\alpha_\pm}
 \end{array}
 \right)	\begin{pmatrix}
 G_{W}^{\pm}	\\
 h_1^{\pm}	\\
  h_2^{\pm}
 \end{pmatrix}. 
 \end{align}
Consequently, we can easily determine $f, \, f_\sigma$ are dependent factors on the free parameters $m_{h_{1,2}^\pm}$, which leads to a convenient numerical investigation of results next, namely
 \begin{equation}
 \label{eq_Higgscouplings}
 f = -\frac{c_{\beta } s_{\beta } \left(2 c_{\alpha_\pm}^2 m_{h_1^\pm}^2 +2 s_{\alpha_\pm}^2 m_{h_2^\pm}^2 -\lambda_{10} v^2\right)}{v_\omega v_\chi}
 \end{equation} 
 will be constrained by the pertubative condition $|f|<4\pi$.
 
Here we determined the relevant factors to sefl-interaction Higgs as follows
\begin{align}\label{factor_hij}
\lambda_{h11}=&v c_{\alpha_{\pm }}^2\left[\lambda_{10} s_{\left(\alpha_0+\beta \right)} + \lambda_5  \left(c_{\beta }^3 s_{\alpha_0}+c_{\alpha_0} s_{\beta }^3\right) + s_{2\beta }  (c_{\beta } c_{\alpha_0} \lambda_1+\lambda_2 s_{\beta } s_{\alpha_0}) \right] 
\crn&+ v s_{\alpha_{\pm }}^2 \left(c_{\beta } s_{\alpha_0}\lambda_{s \rho }+c_{\alpha_0} s_{\beta }\lambda_{s \eta }\right) +\frac{ s^2_{2\alpha_\pm} s_{(\alpha_0+\beta)}(m_{h_1^\pm}^2-m_{h_2^\pm}^2) }{2v},
\crn \lambda_{h22}=&v s_{\alpha_{\pm }}^2 \left[\lambda_{10} s_{\left(\alpha_0+\beta \right)} + \lambda_5 \left(c_{\beta }^3 s_{\alpha_0}+c_{\alpha_0} s_{\beta }^3\right) + s_{2\beta } (c_{\beta } c_{\alpha_0} \lambda_1+\lambda_2 s_{\beta } s_{\alpha_0})\right]  
\crn& + v c_{\alpha_{\pm }}^2  \left(c_{\beta } s_{\alpha_0}\lambda_{s \rho }+c_{\alpha_0} s_{\beta }\lambda_{s \eta }\right) -\frac{ s^2_{2\alpha_{\pm }} s_{\left(\alpha_0+\beta \right)} {(m_{h_1^\pm}^2-m_{h_2^\pm}^2)}}{2v},
\crn \lambda_{h12}=&\lambda_{h21}
\crn=&\frac{vs_{2\alpha_{\pm }}}{2} \left[ \lambda_{10} s_{\left(\alpha_0+\beta \right)} + \lambda_5 \left(c_{\beta }^3 s_{\alpha_0}+c_{\alpha_0} s_{\beta }^3\right) +  s_{2\beta }  (c_{\beta } c_{\alpha_0} \lambda_1+\lambda_2 s_{\beta } s_{\alpha_0}) - \left(c_{\beta } s_{\alpha_0}\lambda_{s \rho }+c_{\alpha_0} s_{\beta }\lambda_{s \eta }\right)\right]
\crn& -\frac{ s_{4\alpha_{\pm }} s_{\left(\alpha_0+\beta \right)} {(m_{h_1^\pm}^2-m_{h_2^\pm}^2)}}{4v},
\end{align}
where, $s_{\left(\alpha_0+\beta \right)} = \text{sin}{\left(\alpha_0+\beta \right)},\, s_{2\alpha_{\pm }} = \text{sin}{2\alpha_{\pm }}, \,  s_{4\alpha_{\pm }} = \text{sin}{4\alpha_{\pm }}$. {We emphasize here that large contributions to $\lambda_{h_{ij}}$ are proportional to ${(m_{h_1^\pm}^2-m_{h_2^\pm}^2)}$. Therefore,  for simplicity in numerical investigation we fix independent Higgs self-couplings to be zero.}

 \section{ \label{app_SMlikeHiggs} The SM-like Higgs boson}
Considering the CP-even scalars, there are three sub-matrices include two $2\times2$ and $4\times4$ for masses of these Higgs bosons in three real (Re) bases $\left(\chi_1^0, \eta_4^0 \right)$, $\left(\phi_2^0, \rho_3^0 \right)$, and $\left(\rho_2^0, \eta_1^0, \phi_3^0, \chi_4^0 \right)$, respectively. Namely
\begin{align}
M^2_{0,14}=&
\left(
\begin{array}{cc}
 \frac{s_{\beta } v^2 (\lambda_{14} s_{\beta }v_\chi -f c_{\beta }  v_\om)}{2 v_\chi} & \frac{1}{2} v (\lambda_{14} s_{\beta }v_\chi-f c_{\beta }  v_\om) \\
 \frac{1}{2} v (\lambda_{14} s_{\beta }v_\chi -f c_{\beta }  v_\om) & \frac{v_\chi(\lambda_{14} s_{\beta } v_\chi-f c_{\beta } v_\om)}{2 s_{\beta }} \\
\end{array}
\right), \\
M^2_{0,23}=&
\left(
\begin{array}{cc}
 \frac{c_{\beta } v^2 (c_{\beta } \lambda_{11} w-f s_{\beta } v_\chi)}{2 v_\om} & \frac{1}{2} (c_{\beta } \lambda_{11} v v_\om-f s_{\beta } v v_\chi) \\
 \frac{1}{2} (c_{\beta } \lambda_{11} v v_\om-f s_{\beta } v v_\chi) & \frac{v_\om (c_{\beta } \lambda_{11} v_\om-f s_{\beta } v_\chi)}{2 c_{\beta }} \\
\end{array}
\right),\\
M^2_{0,44}=&
\left(
\begin{array}{cccc}
 2 c_{\beta }^2 \lambda_2 v^2-\frac{f t_{\beta } V w}{2} & \frac{f V w}{2}+c_{\beta } \lambda_5 s_{\beta } v^2 & \frac{1}{2} f s_{\beta } v V+c_{\beta }\lambda_8 v w & \frac{1}{2} f s_{\beta } v w+c_{\beta } \lambda_9 v V \\
 \frac{f V w}{2}+c_{\beta } \lambda_5 s_{\beta } v^2 & 2 \lambda_1 s_{\beta }^2 v^2 -\frac{f t^{-1}_{\beta } V w}{2} & \frac{1}{2} f c_{\beta } v V+\lambda_6s_{\beta } v w & \frac{1}{2} f c_{\beta } v w+\lambda_7 s_{\beta } v V \\
 \frac{1}{2} f s_{\beta } v V+c_{\beta } \lambda_8 v w & \frac{1}{2} f c_{\beta } v V+\lambda_6 s_{\beta } v w & 2 \lambda_3 w^2-\frac{f c_{\beta } s_{\beta } v^2V}{2 w} & \frac{1}{2} f c_{\beta } s_{\beta } v^2+\lambda '_9 V w \\
 \frac{1}{2} f s_{\beta } v w+c_{\beta } \lambda_9 v V & \frac{1}{2} f c_{\beta } v w+\lambda_7 s_{\beta } v V & \frac{1}{2} f c_{\beta } s_{\beta } v^2+\lambda '_9 V w & 2 \lambda_4 V^2-\frac{f c_{\beta } s_{\beta } v^2 w}{2 V} \\
\end{array}
\right),
\end{align}
where, with each the square matrices $M^2_{0,14}$, and $M^2_{0,23}$ has one zero value are non-hermitian Goldstone bosons $G^0_{14}, \, G^0_{23}$; and $m^2_{H_1^0} = \left(v_1^2+v_\chi^2\right) \left(\frac{\lambda_{14}}{2}-\frac{f v_\om}{2t_{\beta } v_\chi}\right), \, m^2_{H_2^0} = \left(v_2^2+v_\om^2\right) \left(\frac{\lambda_{11}}{2}-\frac{f t_{\beta }v_\chi}{2 v_\om}\right)$ are heavy neutral bosons with masses at two breaking scale of this model, respectively. On the other hand, we can see that $\text{Det}[M^2_{0,44}]\neq0$, but $\text{Det}[M^2_{0,44}]|_{v=0} =0$, which implies that there exists at least one boson with mass on the electroweak scale that can be identified as a Higgs with properties similar to the particle predicted by SM, also known as the SM-like Higgs. Besides, when $\text{Det}[M^2_{0,44}]|_{{v_\omega}=v=0} =0$, the $[M^2_{0,44}] = \text{diag}\left(0,0,0,2\lambda_4v_\chi^2\right)$ which mean that has a very heavy neutral Higgs bosons with mass at the order scale $\mathcal{O}(v_\chi^2)$. As a result, this model can be broken by SSB follows
\begin{equation*}
SU(4)_L \times U(1)_X \xrightarrow{v_\chi} SU(3)_L \times U(1)_N \xrightarrow{v_\om} SU(2)_L \times U(1)_Y \xrightarrow{v} U(1)_Q
\end{equation*}
In particular, detailed calculations are shown as follows
\begin{equation}
C^h_{44}M^2_{0,44}|_{v=0}C^{hT}_{44} = \text{diag}\left(0,-\frac{f v_\chi v_\om}{2 c_{\beta } s_{\beta }},\, \lambda_3 v_\om^2+\lambda_4 v_\chi^2 -\sqrt{\Delta_{44}},\, \lambda_3 v_\om^2+\lambda_4 v_\chi^2 +\sqrt{\Delta_{44}} \right)
\end{equation}
where, $\Delta_{44} = \left(\lambda_3 v_\om^2-\lambda_4 v_\chi^2\right)^2 +{\lambda'}_9^2 v_\chi^2 v_\om^2$, and
\begin{equation}
C^h_{44} = \left(
\begin{array}{cccc}
  c_\beta & s_\beta & 0  & 0 \\
 -s_\beta & c_\beta & 0  & 0 \\
 0 & 0 & c_\xi & -s_\xi \\
 0 & 0 & s_\xi & c_\xi \\
\end{array}
\right), \, t_{2\xi} \equiv \tan(2\xi) = \frac{\lambda'_9 v_\chi v_\om}{\lambda_4 v_\chi^2-\lambda_3 v_\om^2}\propto \mathcal{O}(v_\om/v_\chi)\approx 0,\nonumber
\end{equation}
hence $C^h_{44}M^2_{0,44}C^{hT}_{44}\equiv M^{'2}_{0,44}$.  In this work, we do not represent detailed components in the above matrix because not all of them contribute to calculating the amplitude for LFV decays. On the other hand, we can see in the $\left(\text{Re}[\rho_2^0], \, \text{Re}[\eta_1^0]\right)$ bases the matrix $M^2_{0,44} \to M^2_{h,H}$ which has similar form in Ref.~\cite{Hue:2021zyw} satisfying below
\begin{align}
\left(M^2_{h,H}\right)_{11} &= 2  \left(s_{\beta }^4\lambda_1+ c_{\beta }^4 \lambda_2+c_{\beta }^2 s_{\beta }^2 \lambda_5\right)v^2,\crn
\left(M^2_{h,H}\right)_{22} &= 2 c_{\beta }^2 s_{\beta }^2 (\lambda_1+\lambda_2-\lambda_5)v^2 -\frac{f v_\om v_\chi}{2 c_{\beta } s_{\beta }},\crn
\left(M^2_{h,H}\right)_{12} &= \left(M^2_{h,H}\right)_{21} = c_{\beta } s_{\beta } \left[s_{\beta }^2 (2 \lambda_1-\lambda_5) -c_{\beta }^2 (2 \lambda_2 -\lambda_5)\right] v^2.
\end{align}
Additionally, following the Eqs~.(A1, A2) in previous work in \cite{Hue:2021zyw}, we can define the SM-like Higgs boson as $h$ corresponding to masses $m^2_h \propto \mathcal{O}(v^2)$ \cite{Hue:2021zyw}. Denoting the mixing parameter $\alpha_0$ between the above Higgs bosons and the original states can be written as follows
\begin{align}
\label{ph_stateh}
&\begin{pmatrix}
\sqrt{2}\mathrm{Re}[\rho^0_2] \\
\sqrt{2} \mathrm{Re}[\eta^0_1] 
\end{pmatrix}= \begin{pmatrix}
s_{\alpha_0}& c_{\alpha_0} \\
c_{\alpha_0}&-s_{\alpha_0} 
\end{pmatrix}
\begin{pmatrix}
h \\
H
\end{pmatrix} \equiv U_{(\alpha_0)} M^2_{h,H}U^T_{(\alpha_0)}=\mathrm{diag}\left( m_h^2,\;m_H^2\right)
\crn \to&  \sqrt{2}\mathrm{Re}[\rho^0_2] =  s_{\alpha_0} h + c_{\alpha_0} H, \;\sqrt{2} \mathrm{Re}[\eta^0_1] =  c_{\alpha_0} h - s_{\alpha_0} H.
\end{align}
Identifying with two steps of diagonalizing the Higgs masses shown above, we have
\begin{align}
U_{\alpha_0}=U_{\delta}U_{\beta}=\begin{pmatrix}
c_{\delta}& s_{\delta} \\
s_{\delta}&-c_{\delta} 
\end{pmatrix}\begin{pmatrix}
c_{\beta}& s_{\beta} \\
-s_{\beta}&c_{\beta} 
\end{pmatrix}=\begin{pmatrix}
c_{(\delta+ \beta)}& s_{(\delta+ \beta)} \\
s_{(\delta+ \beta)}&-c_{(\delta+ \beta)} 
\end{pmatrix},
\end{align}
where $\alpha_0=\frac{\pi}{2}-(\delta+\beta)$ and 
\begin{align}
\label{eq:t2d}
 t_{2\delta} \equiv\tan{(2\delta)} =& \frac{2\left(M^2_{h,H}\right)_{12}}{\left(M^2_{h,H}\right)_{22} -\left(M^2_{h,H}\right)_{11}} \crn
=& -\frac{4 c_{\beta } s_{\beta } v^2 \left[c_{\beta }^2 (2 \lambda_2-\lambda_5)+s_{\beta }^2 (\lambda_5-2 \lambda_1)\right]}{v^2 \left[4 \left(c_{\beta }^4
   \lambda_2-c_{\beta }^2 s_{\beta }^2 (\lambda_1+\lambda_2-2 \lambda_5)+\lambda_1 s_{\beta }^4\right)+\lambda_{10}\right] -2c^2_{\alpha_\pm} m^2_{h_1^\pm} -2s^2_{\alpha_\pm} m^2_{h_2^\pm}} 
\crn \propto& \mathcal{O}(v^2/(v_\om v_\chi))\approx 0,
\end{align}
which is consistent with the SM couplings of the SM-like Higgs boson when $\delta \to 0$. 

{Therefore, so to simplify, we fix $\delta=0$ that leads to $M^2 _{h,H}$ to be the unitary matrix and following relationships:
\begin{align}
\left(M^2_{h,H}\right)_{12} \propto&\, s_{\beta }^2 (2 \lambda _1-\lambda _5) -c_{\beta }^2 (2 \lambda _2 -\lambda _5) = 0 \to \, \lambda_5 = 2\dfrac{c^2_\beta \lambda_2 - s^2_\beta \lambda_1}{c^2_\beta - s^2_\beta},\crn
m^2_h = \left(M^2_{h,H}\right)_{11} =& 2  \left(s_{\beta }^4\lambda _1+ c_{\beta }^4 \lambda _2+c_{\beta }^2 s_{\beta }^2 \lambda _5\right)v^2 \to \, \lambda_2 = \lambda_1 t^4_\beta + \dfrac{(c^2_\beta - s^2_\beta)m^2_h}{2c^4_\beta v^2}. 
\end{align}
{Consequently, we can easily obtain $\lambda_2, \, \lambda_5$ as parameters that depend on the $\lambda_1$ input free parameter.}

\end{document}